\documentclass[letter,twocolumn,showpacs,aps,prd,superscriptaddress,nofootinbib,USenglish]{revtex4-1}
\usepackage{atlasphysics}
\usepackage{mathrsfs}
\usepackage{graphicx}
\usepackage{multirow}
\usepackage{xspace}
\usepackage{amsmath}
\usepackage{morefloats}
\usepackage{bigstrut}
\usepackage[caption=false]{subfig}
\usepackage{dcolumn}
\newcommand*{\PYTHIA}{\textsc{Pythia8}\xspace}
\newcommand*{\GEANT}{\textsc{Geant4}\xspace}
\newcommand*{\MADGRAPH}{\textsc{MadGraph5}\xspace}

\newcommand*{\piv}{\ensuremath{\pi_{\rm{v}}}\xspace}
\newcommand*{\mpiv}{\ensuremath{m_{\pi_{\rm{v}}}}\xspace}
\newcommand*{\mPhi}{\ensuremath{m_{\rm{\varPhi}}}\xspace}

\newcommand{\ECal}{ECal\xspace}
\newcommand{\HCal}{HCal\xspace}

\newcommand{\rois}{RoIs\xspace}
\newcommand{\roi}{RoI\xspace}

\newcommand*{\LogCaloRatio}{\log_{10}(E_{\rm{HAD}}/E_{\rm{EM}})\xspace}

\newcommand{\Endcaps}{endcaps\xspace}
\newcommand{\MuonClusterChannel}{Muon Cluster channel\xspace}
\newcommand{\JetChannel}{Jet+\ensuremath{E_{\text{T}}^{\text{miss}}} channel\xspace}

\newcommand*{\dZero}{\ensuremath{d_{\rm{0}}}\xspace}
\newcommand*{\zZero}{\ensuremath{z_{\rm{0}}}\xspace}

\newcommand*{\nEventsJMETTrig}{$N ^ \mathrm{J\MET tr}\xspace $}
\newcommand*{\nEventsMSTrig}{$N^ \mathrm{1 cl}\xspace$}
\newcommand*{\nEventsTwoMSTrigOneUMBarrel}{$N^\mathrm{2 cl}_\mathrm{ 1 UM B cl}\xspace$}
\newcommand*{\nEventsTwoMSTrigOneUMEndcaps}{$N^\mathrm{2 cl}_\mathrm{ 1 UM E cl}\xspace$}
\newcommand*{\PMSVertex}{$P^\mathrm{vx}_\mathrm{no MS tr}\xspace$}
\newcommand*{\PMSRecoBarrel}{$P^\mathrm{vx}_\mathrm{B cl}\xspace$}
\newcommand*{\PMSRecoEndcaps}{$P^\mathrm{vx}_\mathrm{E cl}\xspace$}

\usepackage[cernpreprint]{atlascover}

\usepackage{hyperref}
  \hypersetup{pdftitle={ATLAS Document},pdfauthor={ATLAS Collaboration}} 
  \hypersetup{colorlinks,breaklinks}
  \hypersetup{linkcolor=blue,citecolor=blue,filecolor=black,urlcolor=blue}

\pacs{12.60.i, 13.85.Rm}

\AtlasTitle{Search for long-lived, weakly interacting particles that decay to displaced hadronic jets in proton--proton collisions at $\sqrt{s}=8$ TeV with the ATLAS detector}

\PreprintIdNumber{CERN-PH-EP-2015-071}  

\AtlasAbstract{A search for the	 decay of neutral, weakly interacting, long-lived particles using data collected by the ATLAS detector at the LHC is presented. This analysis uses the full dataset recorded in 2012: 20.3~fb$^{-1}$ of proton--proton collision data at $\sqrt{s} = 8$~\TeV. The search employs techniques for reconstructing decay vertices of long-lived particles decaying to jets in the inner tracking detector and muon spectrometer. Signal events require at least two reconstructed vertices. No significant excess of events over the expected background is found, and limits as a function of proper lifetime are reported for the decay of the Higgs boson and other scalar bosons to long-lived particles and for Hidden Valley \Zprime and Stealth SUSY benchmark models. The first search results for displaced decays in \Zprime and Stealth SUSY models are presented. The upper bounds of the excluded proper lifetimes are the most stringent to date.}

\AtlasJournal{Phys.~Rev.~D}  

\begin{document}

\title{\AtlasTitleText}
\author{The ATLAS Collaboration}
\begin{abstract}
	\AtlasAbstractText
\end{abstract}
\maketitle
\section{Introduction}\label{sec:intro}

This paper describes a search for long-lived neutral particles (LLPs), produced by proton--proton interactions at $\sqrt{s} = 8$~\TeV, that decay to hadronic jets far from the interaction point (IP). The search is performed using data collected with the ATLAS detector at the Large Hadron Collider (LHC). These decays can result in secondary decay vertices (displaced vertices) that can be highly displaced from the IP. This analysis focuses on displaced vertices occurring in the inner tracking detector (ID) and muon spectrometer (MS).  The event selection criteria and vertex reconstruction algorithms employed by this analysis result in a generic, model-independent selection of candidate events with two displaced vertices.

Many extensions of the Standard Model (SM) include particles that are neutral, weakly coupled, and long-lived  that can decay to final states containing several hadronic jets. Such extensions include gauge-mediated supersymmetry breaking (GMSB) models~\cite{SUSY}, the minimal supersymmetric standard model (MSSM) with R-parity violation \cite{MSSM}, inelastic dark matter models~\cite{DarkMatter}, Hidden Valley scenarios~\cite{HV1,HV2,HV3}, stealth supersymmetry (Stealth SUSY) models~\cite{Josh1,Josh2}, two-Higgs-doublet models (2HDM)~\cite{2HDM}, and a recent baryogenesis model~\cite{Baryogenesis}. Independent of a specific model, Higgs fields (scalar fields) can couple to hidden-sector fields that are singlets under the SM gauge group.  Such couplings can result in the SM sector mixing with a hidden sector~\cite{PandW} and depending on the value of these couplings, the lifetime of the hidden-sector particles can be sufficiently long to result in highly displaced decays. 

The results of this search are interpreted in the context of three benchmark models. Two models are Hidden Valley scenarios. The first involves the decay of a scalar boson to a pair of long-lived neutral particles (scalars or pseudoscalars) that each decay to a pair of hadronic jets. 
The second model has a heavy \Zprime boson that decays to long-lived, as well as prompt, neutral particles that decay to hadronic jets. The third is a variant of Stealth SUSY in which a pair of gluinos decay to long-lived singlinos, which in turn each decay to a low-mass gravitino and a pair of hadronic jets. While all measurements of branching ratios and couplings for the recently discovered Higgs boson are consistent with SM expectations~\cite{Coupling1,Coupling2,Coupling3,CMSCoupling}, the current measurement uncertainties do not exclude non-SM decay modes with a branching ratio of up to 30$\%$~\cite{CMSCoupling,NonSMCoupling1,NonSMCoupling2}. 
Therefore, the decay of the Higgs boson to a pair of long-lived scalars or pseudoscalars is also explored. Such long-lived scalars or pseudoscalars that couple to the SM by mixing with a scalar boson or the Higgs boson do so via a Yukawa coupling, and thus the couplings to fermions and antifermions are proportional to the fermion mass.

To date, no search has observed evidence of long-lived neutral particles decaying to hadronic jets. Searches were carried out at the Tevatron (\mbox{$\sqrt{s}=1.96\TeV$}) by both the CDF~\cite{CDF} and D0~\cite{D0} collaborations, at the LHC by the LHCb collaboration in proton--proton collisions at \mbox{$\sqrt{s}=7\TeV$}~\cite{LHCb}, and more recently by the ATLAS and CMS collaborations~\cite{SUSYDV,CMS} at \mbox{$\sqrt{s}=8\TeV$}. 
These searches were mainly restricted to decays within the inner tracking volume of the detectors and thus explored primarily a mean proper lifetime ($c\tau$) of less than 1~m.  

This work significantly extends the $c\tau$ range of the ATLAS search for a light scalar boson decaying to long-lived neutral particles at $\sqrt{s}=7\TeV$ in 1.94~fb$^{-1}$ of 2011 proton--proton collision data at the LHC~\cite{HVAtlas} that covered the $c\tau$ region 1 to 20~m.  Additionally, it extends the range of proper lifetimes excluded by a recent ATLAS analysis~\cite{CalRatio}, which uses the same scalar boson model and mass points but focuses on displaced decays in the hadronic calorimeter. 

This analysis also reports the results of the first search for displaced hadronic jets from heavy \Zprime boson and Stealth SUSY decays.


\section{ATLAS detector}\label{sec:ATLASdetector}
The ATLAS detector~\cite{atlas-detector}, which has nearly 4$\pi$ steradian coverage in solid angle, is a multipurpose detector consisting of an inner tracking detector embedded in a                                                                                                                                                                           superconducting solenoid, electromagnetic and hadronic calorimeters (ECal and HCal), and a muon spectrometer incorporating magnetic fields produced by three superconducting, air-core toroidal magnets each comprised of eight coils. The ID covers the range 0.05~m~$<r<$1.1~m and $|z|<3.5$~m.\footnote{ATLAS uses a right-handed coordinate system with its origin at the nominal interaction point (IP) in the center of the detector and the $z$-axis along the beam pipe.  The $x$-axis points from the IP to the center of the LHC ring, and the $y$-axis points upwards. Cylindrical coordinates $(r,\phi)$ are used in the transverse plane, where $\phi$ is the azimuthal angle around the $z$-axis. The pseudorapidity is defined in terms of the polar angle $\theta$ as $\eta = - \ln \tan(\theta/2)$. Angular distance is measured in units of $\Delta R \equiv \sqrt{(\Delta\eta)^{2} + (\Delta\phi)^{2}}$.}  It consists of a silicon pixel detector, a silicon microstrip detector (semiconductor tracker, SCT), and a straw tube tracker (transition radiation tracker, TRT).  Together, the three systems provide precision tracking of charged particles for \mbox{$|\eta|<2.5$}. Three radial layers of high-granularity silicon pixel detectors cover the region around the IP followed by the SCT, which provides up to four two-dimensional measurement points per track. The TRT provides additional information for track reconstruction in the region $|\eta|<2.0$.

The calorimeter system covers the pseudorapidity range $|\eta| < 4.9$. Within the region $|\eta|< 3.2$, electromagnetic calorimetry is provided by high-granularity, liquid-argon (LAr) barrel and endcap electromagnetic calorimeters with lead absorbers. An additional thin LAr presampler covering $|\eta| < 1.8$ is used to correct for energy loss in material upstream of the calorimeters. The \ECal extends from 1.5\,m to 2.0\,m in $r$ in the barrel and from 3.6\,m to 4.25\,m in $|z|$ in the \Endcaps. 

Hadronic calorimetry is provided by a steel/scintillating-tile calorimeter, segmented into three barrel structures within $|\eta| < 1.7$, and two copper/LAr hadronic endcap calorimeters. The solid angle coverage is completed with forward copper/LAr and tungsten/LAr calorimeter modules optimized for electromagnetic and hadronic measurements, respectively. The \HCal covers the region from 2.25\,m to 4.25\,m in $r$ in the barrel and from 4.3\,m to 6.05\,m in $|z|$ in the \Endcaps. Together the ECal and HCal have a thickness of 9.7 interaction lengths at $\eta=0$.

The MS comprises separate trigger and tracking chambers that measure the deflection of muons in a magnetic field generated by superconducting air-core toroids. The chamber system covers the region $|\eta| < 2.7$ with three layers of monitored drift tubes, complemented by cathode strip chambers in the forward region.

Three stations of resistive plate chambers (RPC) and thin gap chambers (TGC) are used for triggering in the MS barrel and \Endcaps, respectively. The barrel chamber system is subdivided into sixteen sectors: eight large sectors (between the magnet coils) and eight small sectors (inside the magnet coils). As a result, the first two RPC stations, which are radially separated by 0.5~m, start at a radius of either 7\,m (large sectors) or 8\,m (small sectors). The third station is located at a radius of either 9\,m (large sectors) or 10\,m (small sectors). In the \Endcaps the first TGC station is located at $|z|=13$~m. The other two stations start at $|z|=14$~m and are separated by $0.5$~m in $z$.  The muon trigger system covers the range $|\eta| < 2.4$. 

The ATLAS trigger system has three levels~\cite{TRIG}.  The first level (L1), which is a hardware-based system, uses coarse data collected from the calorimeters and muon detectors. The second (L2) and third (Event Filter, EF) levels are software-based systems that use information from all of the ATLAS subdetectors. Together, L2 and EF are called the High Level Trigger (HLT). 
L1 thresholds are applied to transverse energy ($\ET$) for calorimeter triggers and transverse momentum ($\pT$) for muon triggers. The L1 trigger identifies ($\eta$,$\phi$)  Regions of Interest (\rois) associated with a specific physics signature. The HLT makes use of the \rois to access the full information for the most important parts of the event.

The implementation of the L1 muon trigger logic is similar for both the RPC and TGC systems. Each of the three planes of the RPC system and the two outermost planes of the TGC system consist of a doublet of independent detector layers.  The first TGC plane contains three detector layers. A low-$\pT$ muon \roi is generated by requiring a coincidence of hits in at least three of the four layers of the two inner RPC planes for the barrel and of the two outer TGC planes for the \Endcaps. A high-$\pT$ muon \roi requires, in addition to a low-$\pT$ muon, hits in at least one of the two layers of the outer RPC plane for the barrel or in two of the three layers of the innermost TGC plane for the \Endcaps. The muon \rois have a spatial extent of $0.2\times0.2$ in $\Delta\eta \times \Delta\phi$ in the MS barrel and $0.1\times0.1$ in $\Delta\eta \times \Delta\phi$ in the MS \Endcaps. Only the two highest-$\pT$ \rois per MS sector are used by the HLT.

The L1 calorimeter trigger is based on information from the calorimeter elements within projective regions, called trigger towers. The trigger towers have a size of approximately 0.1 in $\Delta\eta$ and $\Delta\phi$ in the central part of the calorimeter, \mbox{$|\eta|<2.5$}, and are larger and less uniform in the more forward region.

\section{Analysis strategy}\label{sec:strategy}
Hadronic decays that are displaced from the IP leave a unique detector signature that can be reconstructed as a displaced vertex. This analysis searches for events with two displaced vertices in either the ID or MS, or one in each.

This analysis studies two separate channels, defined by the triggers used to select events. The ATLAS Muon RoI Cluster trigger~\cite{MuonRoITrigger} is used to preselect events that satisfy displaced-decay criteria in the MS. The search for both the scalar boson and Stealth SUSY models makes use of events selected by the Muon RoI Cluster trigger. The sample of events selected by this trigger belongs to the \MuonClusterChannel. 

The large multiplicity of long-lived particles in the \Zprime benchmark samples causes many events to fail to satisfy the isolation criteria of the Muon RoI Cluster trigger, but also provides other objects on which to trigger. The large number of long-lived particles per event results in a higher probability for at least one of the particles to decay sufficiently promptly to generate a jet and another to decay with a sufficiently large radius to contribute to missing transverse momentum. The negative vector sum of the momenta in the plane transverse to the beam line of all particles detected in a \textit{pp} collision is called missing transverse momentum, $\boldsymbol{E}_{\rm{T}}^{\rm{miss}}$, and its magnitude is denoted by \MET~\cite{MET,MET8}. 

A jet plus \MET trigger is thus used for selecting events from the \Zprime model. It allows for the inclusion of events with displaced decays in the ID but none in the MS. These events belong to the Jet+\MET channel.

In scalar boson and Stealth SUSY events the two displaced decays are in an approximately back-to-back configuration and thus events do not have high \MET. Additionally, unless one of the LLPs decays in the ID, scalar boson benchmark sample events would not contain a sufficiently high energy jet to satisfy the jet $\pT$ requirement of the Jet+\MET trigger.

To reconstruct the vertex of a displaced decay, the analysis uses two separate reconstruction algorithms: one for inner detector vertices (described in Sec.~\ref{sec:dispvrt}) and one for muon spectrometer vertices~\cite{MSVXRECO}.  In order to reduce backgrounds, an event is considered as a signal candidate only if two displaced decays are reconstructed. This selection results in the following event topologies: two ID vertices (2IDVx), two MS vertices (2MSVx) and one vertex in the MS and one in the ID (IDVx+MSVx). The analysis is structured in a way that makes it sensitive to many models with displaced decays. Although events are required to have two reconstructed vertices, other reconstructed objects may be present. The topologies applicable and models analysed for each analysis channel are summarized in Table~\ref{tab:topo}.
 
\begin{table}[hbtp]
\caption{The topologies considered in the analysis and the corresponding triggers and benchmark models.}
\begin{ruledtabular}
\begin{tabular}{lll}
\multicolumn{1}{c}{Trigger} &\multicolumn{1}{c}{Applicable topologies} & \multicolumn{1}{c}{Benchmarks} \\
\hline
Muon RoI & \multirow{2}{*}{IDVx+MSVx, 2MSVx} &  Scalar boson, \bigstrut[t] \\
  Cluster & & Stealth SUSY \bigstrut[b] \\
Jet+\MET & 2IDVx, IDVx+MSVx, 2MSVx & \Zprime \bigstrut \\
\end{tabular}
\end{ruledtabular}
\label{tab:topo}
\end{table}	

\section{Description of benchmark models}\label{sec:models}
\begin{figure*}[thbp]
\centering
\begin{tabular}{ccc}
 \subfloat[]{\raisebox{3.5 mm}{\includegraphics[width=0.3\textwidth]{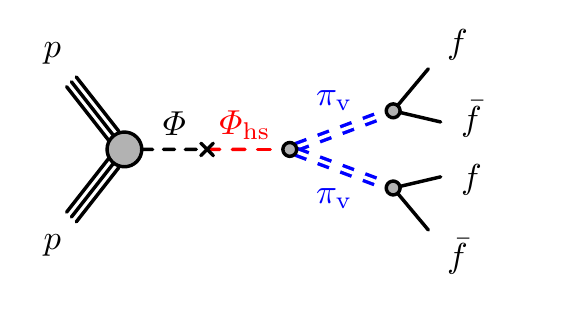}\label{fig:HiggsFeyn}}} &
 \subfloat[]{\includegraphics[width=0.3\textwidth]{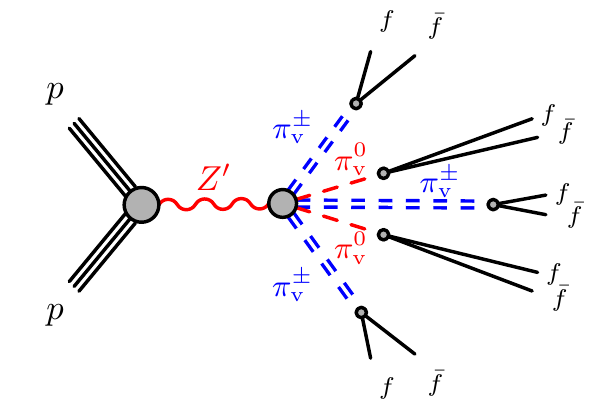}\label{fig:ZprimeFeyn}} &
 \subfloat[]{\raisebox{0.5 mm}{{\includegraphics[width=0.3\textwidth]{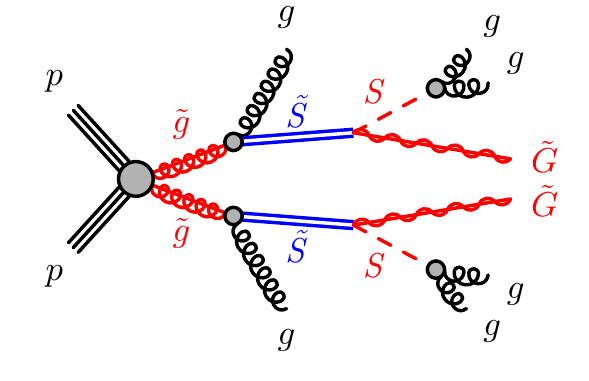}\label{fig:StealthFeyn}}}} \\
 \end{tabular}
\caption{
Diagrams of the benchmark models studied in this analysis: (a) the Higgs boson or scalar boson model, (b) the Hidden Valley \Zprime model, and (c) the Stealth SUSY model. The long-lived particles in these processes are represented by double lines and labeled (a) $\piv$, (b) $\piv^{\pm}$, and (c) \textit{\~{S}}. The gravitino, \textit{\~{G}}, has very low mass and does not carry away a substantial amount of energy.}
\label{fig:Diagrams}
\end{figure*}

\subsection{Hidden Valley}
Hidden Valley (HV) scenarios~\cite{HV1,HV2,HV3} are a general class of models in which a hidden sector (v-sector) is added to the SM. All of the SM particles are neutral under the v-sector gauge group, and the v-sector particles (v-particles) are neutral under the SM gauge group. Because the v-particles do not couple directly to the SM, decays of v-particles to SM particles must occur via communicator particles. Depending on the masses of the communicators and their couplings, this can result in long lifetimes for the lightest v-particles. Two communicators are considered, giving rise to different signatures in the detector. 

The first scenario, shown in Fig.~\ref{fig:Diagrams}(a), predicts a scalar boson $\varPhi$ mixing with a hidden-sector scalar boson $\varPhi_{\rm{hs}}$. The hidden sector contains the communicator, $\varPhi_{\rm{hs}}$, and a v-isospin triplet of \piv particles. The $\varPhi_{\rm{hs}}$ decays to a pair of these \piv particles, which then decay back to SM particles, again via $\varPhi_{\rm{hs}}$--$\varPhi$ mixing. Because of the Yukawa coupling, each \piv decays predominantly to heavy fermions, $b\overline{b}$, $c\overline{c}$, and $\tau^{+}\tau^{-}$ in the ratio 85:5:8, provided $\mpiv<2m_{\rm{top}}$. 
The branching ratio for $\varPhi$ decaying into a pair of hidden sector particles is not constrained in these models. It is therefore interesting to focus both on Higgs boson decays to long-lived particles, where $\varPhi$ is the SM Higgs boson, and on other $\varPhi$ mass regions previously unexplored for decays to long-lived particles. 

The second scenario involves a massive communicator (\Zprime) produced by quark--antiquark annihilation and decaying into the hidden sector via $\Zprime\rightarrow q_{\rm{v}}\bar{q}_{\rm{v}}$. The v-quarks hadronize into showers of \piv particles. For the particular mass benchmarks used in this analysis, the $\pi^{0}_{\rm{v}}$ decays promptly, while the $\pi^{\pm}_{\rm{v}}$ lifetime is a free parameter. The $\pm$ and $0$ superscripts for the \piv indicate a charge under the v-sector gauge group, v-isospin, and not electric charge. This process is shown in Fig.~\ref{fig:Diagrams}(b).

\subsection{Stealth SUSY}
Stealth SUSY models~\cite{Josh1,Josh2} are a class of R-parity-conserving SUSY models that do not have large \MET signatures. While this can be accomplished in many different ways, this search explores a model that involves adding a hidden-sector (stealth) singlet superfield $S$ at the electroweak scale, which has a superpartner singlino $\tilde{S}$. By weakly coupling the hidden sector to the MSSM, the mass splitting between $S$ and $\tilde{S}$ ($\delta M$) is small, assuming low-scale SUSY breaking. High-scale SUSY breaking also can be consistent with small mass splitting and Stealth SUSY, although this requires a more complex model and is not considered in this search~\cite{Josh2}.

The SUSY decay chain ends with the singlino decaying to a singlet plus a low-mass gravitino $\tilde{G}$, where the gravitino carries off very little energy and the singlet promptly decays to two gluons. The effective decay processes are $\tilde{g} \rightarrow \tilde{S} g$ (prompt), $\tilde{S} \rightarrow S \tilde{G}$ (not prompt), and $S \rightarrow g g$ (prompt). This scenario results in one prompt gluon and two displaced gluon jets per gluino decay. Since R-parity is conserved, each event necessarily produces two gluinos, resulting in two displaced vertices. A Feynman diagram of this process is shown in Fig.~\ref{fig:Diagrams}(c).

The decay width (and, consequently, the lifetime) of the singlino is determined by both the mass splitting $\delta M$ and the SUSY-breaking scale $\sqrt{F}$: ${\Gamma}_{\tilde{S}~\rightarrow~S \tilde{G}}~\approx~m_{\tilde{S}}(\delta M)^4/\pi F^2$ ~\cite{Josh1}. The SUSY-breaking scale $\sqrt{F}$ is not a fixed parameter, and thus the singlino has the possibility of traveling an appreciable distance through the detector, leading to a significantly displaced vertex. 

\section{Data and simulation samples}\label{sec:samples}
The dataset used in this analysis was recorded by the ATLAS detector in the 2012 run during periods in which all subdetectors relevant to the analysis were operating efficiently. The integrated luminosity is 20.3 fb$^{-1}$. The \MuonClusterChannel uses only 19.5 fb$^{-1}$ of data because the trigger was not active at the beginning of the run. The uncertainty on the integrated luminosity, estimated following the methodology described in Ref.~\cite{Lumi2011}, is 2.8\%. 

Monte Carlo (MC) simulation samples are produced for scalar boson, \Zprime, and Stealth SUSY models. The masses are chosen to span the accessible parameter space, and the proper lifetime values are chosen to maximize the distribution of decays throughout the ATLAS detector volume. The masses for each sample are listed in Table~\ref{tab:MCSamples}. Approximately 400,000 events are generated for each sample.

For the scalar boson model, the MC simulation events are generated with \PYTHIA~\cite{PYTHIA8}. The \Zprime events are simulated in a two-step process. An external generator, HVMC, is used to simulate \Zprime production and decay to hidden-sector particles~\cite{HVMC}. The decay of \piv to SM particles and their subsequent decays, as well as the showering and hadronization of SM partons, are simulated using \PYTHIA. Stealth SUSY events are generated with \MADGRAPH~\cite{Madgraph5}, and \PYTHIA is used for hadronization. 
The parameterization used for the proton parton distribution function (PDF) for the scalar boson and \Zprime simulations is MSTW2008~\cite{MSTW2008}, while CTEQ 6L1~\cite{CTEQ6L1} is used for Stealth SUSY. 

\begin{table}[h!]
\begin{ruledtabular}
\caption{\label{tab:MCSamples}Mass parameters for the simulated scalar boson, \Zprime and Stealth SUSY models.}
\begin{tabular}{cl}
Scalar boson mass [GeV] & \multicolumn{1}{c}{\piv mass [GeV]} \\
\hline
100 & 10, 25 \bigstrut[t] \\
125 & 10, 25, 40 \\
140 & 10, 20, 40 \\
300 & 50 \\
600 & 50, 150 \\
900 & 50, 150 \bigstrut[b] \\
\hline\hline
\Zprime mass [TeV] & \multicolumn{1}{c}{\piv mass [GeV]} \bigstrut \\
\hline
1 & 50 \bigstrut[t] \\
2 & 50 \\
2 & 120 \bigstrut[b] \\
\hline\hline
$\tilde{g}$ mass [GeV] & \multicolumn{1}{c}{$\tilde{S}$, $S$ mass [GeV]}\bigstrut \\
\hline
110 & 100, 90 \bigstrut[t] \\
250 & 100, 90 \\
500 & 100, 90 \\
800 & 100, 90 \\
1200 & 100, 90 \\
\end{tabular}
\end{ruledtabular}
\end{table}

For all simulated samples, the propagation of particles through the ATLAS detector is modeled with \GEANT \cite{GEANT4a} using the full ATLAS detector simulation \cite{GEANT4b} for all the simulated samples. In addition, each MC sample is overlaid with zero-bias data events that are selected from bunch crossings corresponding to one full revolution around the LHC after a high-\pT interaction. This overlaid data sample correctly represents all sources of detector background such as cavern background (a gas of thermal neutrons and photons filling the ATLAS cavern during ATLAS operation), beam halo, cosmic rays and electronic noise.  It also correctly reproduces pileup interactions (multiple interactions per bunch crossing).

The analysis employs data-driven techniques to estimate the backgrounds. Two separate samples from data are used: multijet and minimum-bias \cite{minbias} events. The ATLAS minimum bias trigger selects events with activity in the range $2.1 < | \eta | < 3.8$ and $360^\circ$ in $\phi$. Multijet events are selected using a set of single-jet triggers with various transverse energy thresholds.  Simulated multijet samples are employed for characterizing displaced vertices when the multijet data sample is statistically limited.  These are generated with \PYTHIA and the CT10 PDF set \cite{CT10}. 

\section{Trigger} \label{sec:trigger}
\subsection{Muon RoI Cluster trigger}\label{subsec:TriggerMuonRoiCluster}

The Muon RoI Cluster trigger is a signature-driven trigger that selects decays of neutral particles in the MS. It is used to select candidate events for the scalar boson and Stealth SUSY searches, and is efficient for hadronic decays occurring in the region from the outer radius of the HCal to the middle of the MS. The trigger selects events with a cluster of muon RoIs contained in a $\Delta R=0.4$ cone that are preceded by little or no activity in the ID or calorimeters. This isolation requirement reduces backgrounds from muon bremsstrahlung and punch-through jets. A punch-through jet is a hadronic or electromagnetic shower not contained in the calorimeter volume, which results in tracks in the MS. The details of the performance and implementation of this trigger on both a set of MC simulated benchmark samples and data can be found in Ref.~\cite{MuonRoITrigger}. 

\begin{figure}[htbp]
\includegraphics[width=0.5\textwidth]{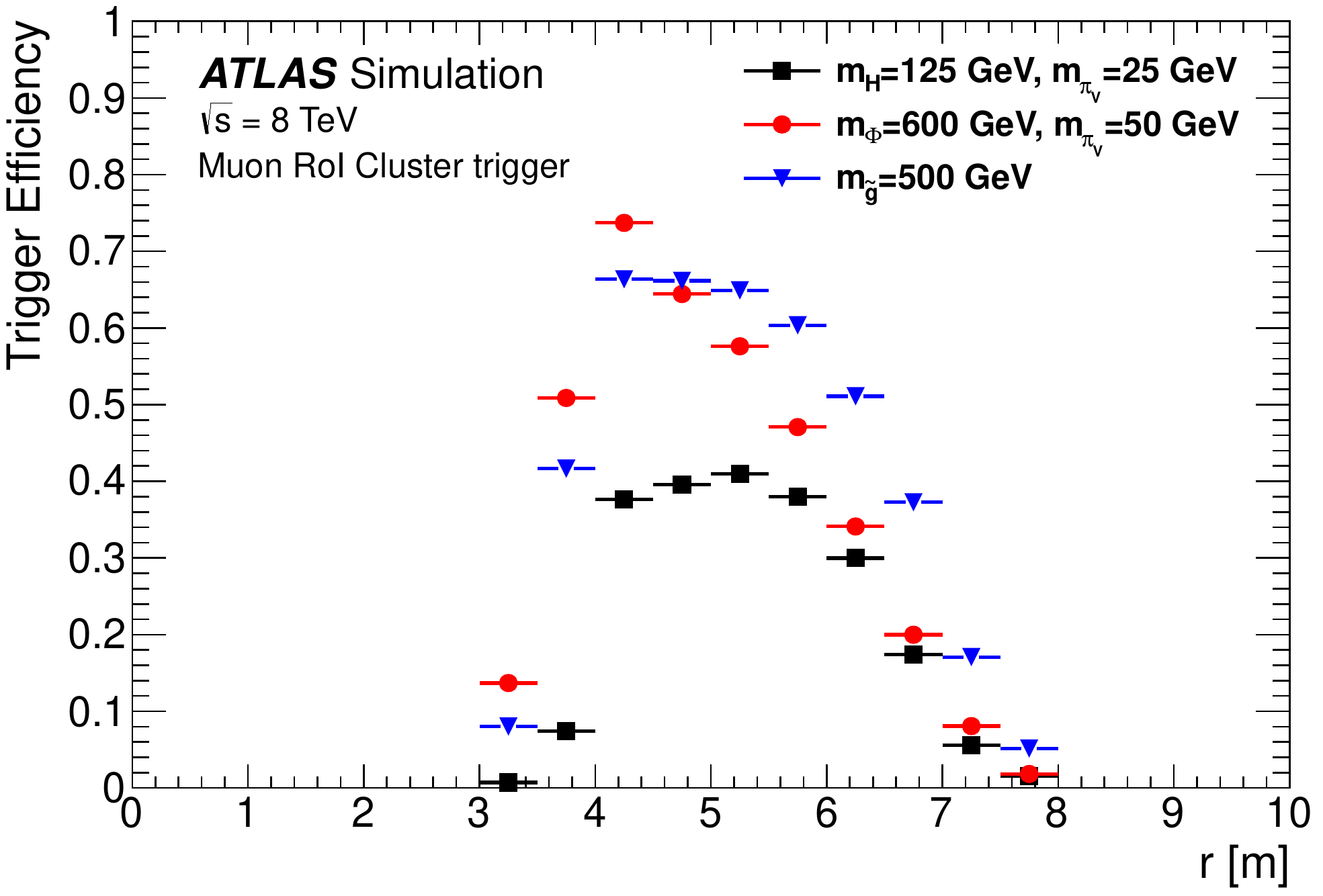}
\caption{Efficiency for the Muon RoI Cluster trigger in the barrel as a function of the decay position of the long-lived particle for three simulated benchmark samples.}
\label{fig:MuonClusterTriggerEff-Barrel}
\end{figure}

\begin{figure}[htbp]
\includegraphics[width=0.5\textwidth]{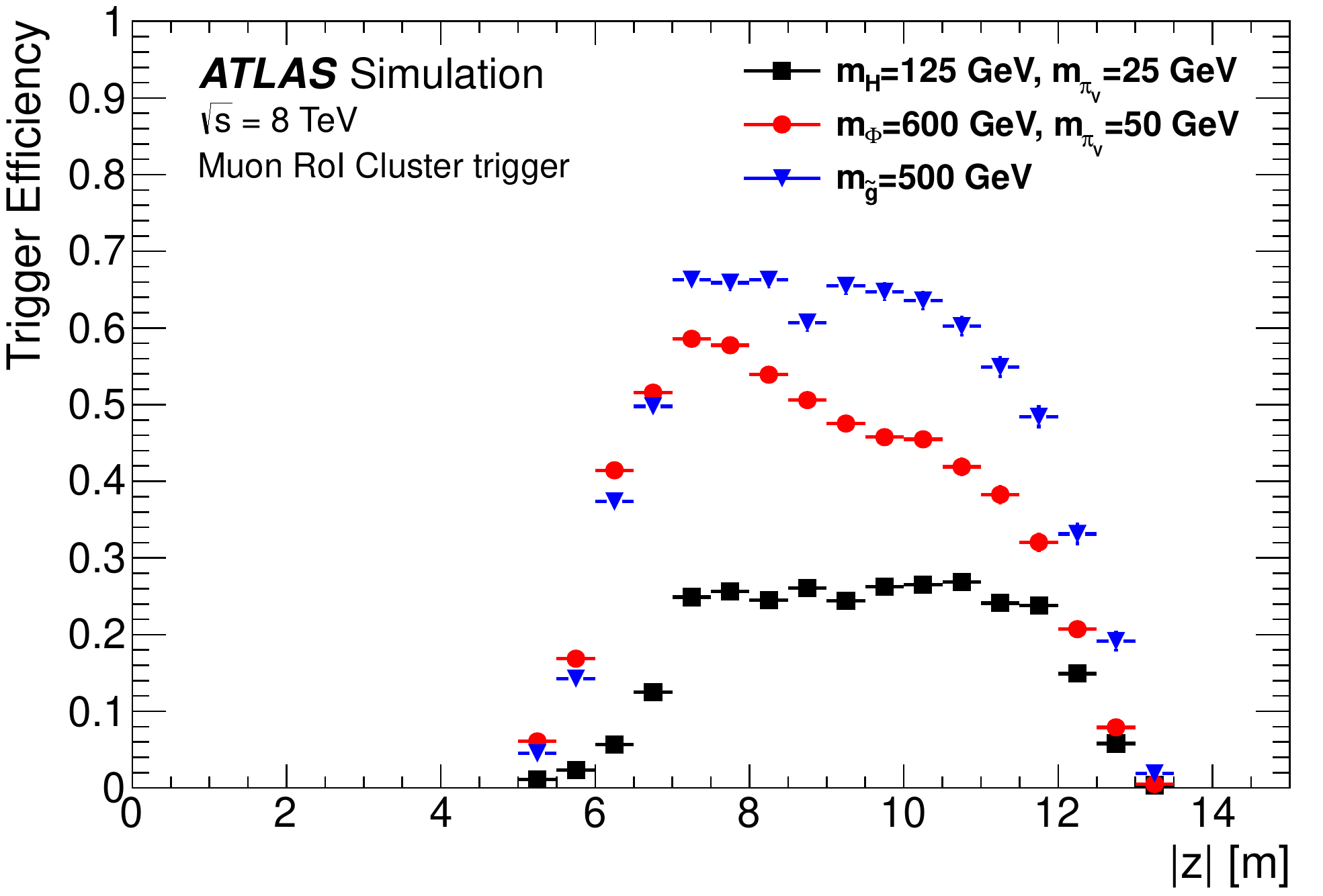}
\caption{Efficiency for the Muon RoI Cluster trigger in the \Endcaps as a function of the the decay position of the long-lived particle for three simulated benchmark samples.}
\label{fig:MuonClusterTriggerEff-Endcaps}
\end{figure}

The trigger efficiency, defined as the fraction of long-lived particles selected by the trigger as a function of the long-lived particle decay position, is shown in Figs.~\ref{fig:MuonClusterTriggerEff-Barrel}~and~\ref{fig:MuonClusterTriggerEff-Endcaps} for three MC simulated benchmark samples. The uncertainties shown are statistical only. The relative differences in efficiencies are a result of the different kinematics between benchmark samples.

\subsection{Jet+\MET trigger}\label{subsec:TriggerZPrime}
A single-jet plus \MET trigger is employed for the \Zprime search. The trigger uses a leading jet \eT threshold of 110 GeV and an \MET threshold of 75~\GeV. Offline requirements are a leading jet $\pT\geq120$~\GeV and $\MET \geq 200$~\GeV, which result in a constant trigger efficiency as a function of both $\pT$ and \MET. For events passing the offline requirements, the trigger efficiency is 87--100\%, depending on the MC simulated \Zprime benchmark sample.

\section{Physics objects and displaced vertex reconstruction} \label{sec:dispvrt}
Hadronic jets, \MET, and displaced vertices are used in this search.

The anti-$k_t$ jet algorithm~\cite{AntiKt} with a radius parameter of 0.4 is used to reconstruct jets from topological clusters, which are three-dimensional clusters of neighboring energy deposits in the calorimeter cells.  A calibration procedure is used in which the raw energy measurements from these clusters are corrected to the jet energy scale~\cite{JETPERF2}. Identifying displaced long-lived particle decays in the ATLAS detector critically depends on the ability to reliably reconstruct displaced vertices.  In the ATLAS MS, the standalone MS vertex reconstruction algorithm~\cite{MSVXRECO} is used to reconstruct displaced vertices. Displaced vertices in the ID are reconstructed using a modified version of the algorithm used for reconstructing primary vertices~\cite{PriVrtReco}.

\subsection{Track and vertex reconstruction in the inner tracking detector}

The default ID track reconstruction algorithm~\cite{PriVrtReco} uses hits in the silicon detectors to form tracks, and then adds TRT hits to extend the tracks.  Next, the algorithm constructs segments from TRT hits and extrapolates to add in leftover silicon hits (pixel and SCT hits).  After the default track reconstruction is complete, there remains a collection of hits that are not associated with any track. The signatures studied in this analysis produce many such unassociated hits because the impact parameters of tracks from a displaced decay are often larger than the maximum impact parameter allowed by the default track reconstruction algorithm. During track reconstruction, the impact parameters are calculated as the distance of a track from the nominal detector center.

Displaced tracks are reconstructed by using the unassociated hits in a second iteration of track finding. This second iteration is performed with modified requirements on the track impact parameters, the minimum number of silicon hits per track that are not shared with another track, and the maximum number of hits shared with another track. Displaced vertices from decays that occur in the outermost region of the pixel detector or just before silicon layers are more likely to have a high fraction of hits shared with another track, since the decay products may not be well separated. The default and modified requirements are listed in the upper part of Table~\ref{tab:IDvertexParam}, where the transverse and longitudinal impact parameters are denoted by \dZero and \zZero, respectively. 

An algorithm to identify secondary vertices in the ID was developed for this analysis. The algorithm, based on the ATLAS primary vertex reconstruction algorithm, searches for a group of tracks and then performs a fit to find candidate vertices. The constraint that the reconstructed vertex must be compatible with the beam spot position is removed, and the parameters of the tracks usable for vertex finding are modified. Here, impact parameters are recalculated to be with respect to the center of the beam spot. The lower part of Table~\ref{tab:IDvertexParam} summarizes the changes from the default selection made to the track impact parameter significances and the number of silicon hits required per track. In addition, a minimum \dZero requirement of $10~$mm is applied, which removes any remaining tracks originating from the IP. If a reconstructed vertex has associated tracks that have hits at radii smaller than the radial distance of the vertex from the beamline, those tracks are removed and the vertex is refitted.

\begin{table}

\caption{Default and modified parameters used for track reconstruction and for the selection of tracks used for displaced vertex reconstruction.
}
\begin{ruledtabular}
\begin{tabular}{lcc}
\multicolumn{3}{c}{Track reconstruction}\\
\multicolumn{1}{c}{Parameter} & Default value & Modified value \bigstrut \\
\hline
Max. \dZero & 10\,mm & 500\,mm \bigstrut[t] \\
Max. $|z_0|$ & 320\,mm & 1000\,mm \\
Min. number of silicon hits & 6 & 2 \\
Max. number of shared hits & 1 & 2 \bigstrut[b] \\
\hline\hline
\multicolumn{3}{c}{Tracks for vertex reconstruction} \bigstrut \\
\multicolumn{1}{c}{Parameter} & Default value & Modified value \bigstrut \\
\hline
Min. \dZero & -- & 10 mm \bigstrut[t] \\
Max. \dZero/$\sigma(\dZero)$ & 5 & -- \\
Max. $|z_0|$/$\sigma(\zZero)$ & 10 & -- \\
Min. number of silicon hits & 6 & 4 \\
Min. number of pixel hits & 1 & 0 \\
Min. number of SCT hits & 4 & 2 \\
Max. track $\chi^{2}/$d.\,o.\,f. & 3.5 & 5\\
\end{tabular}
\end{ruledtabular}
\label{tab:IDvertexParam}
\end{table}

In MC simulated benchmark samples, reconstructed vertices are considered matched to simulated long-lived particle decay vertices if the reconstructed vertex is within 5\,mm of the simulated decay position. In addition, at least two tracks from the reconstructed vertex are required to be matched to particles from the simulated decay vertices.

A significant fraction of background vertices originate from hadronic interactions in material layers of the silicon detectors. To remove this background, a material veto is implemented by using an algorithm that calculates the distance between a vertex and the closest material layer. The methodology is similar to that used in Ref.~\cite{HadInteraction}. 

To remove vertices that are consistent with originating from within a material layer, the variable $d/\sigma$ is defined, where $d/\sigma$ is the distance to the closest material layer divided by the vertex position uncertainty. Vertices with $d/\sigma$ $<$ 6 are assumed to be due to material interactions and are therefore rejected. For vertices in the pixel layers, the vertex position is transformed into the local coordinates of the closest pixel module, and vertices that are reconstructed in the module volume are vetoed. Vertices reconstructed in the pixel support structures in both the barrel and the endcaps are also vetoed. 
	
To reject poorly reconstructed vertices resulting from random track crossings or a prompt track passing close to a displaced low track-multiplicity secondary vertex,  the $\chi^2$ probability of the vertex fit is required to be greater than 0.001.

\begin{figure}[htbp]
\includegraphics[width=0.5\textwidth]{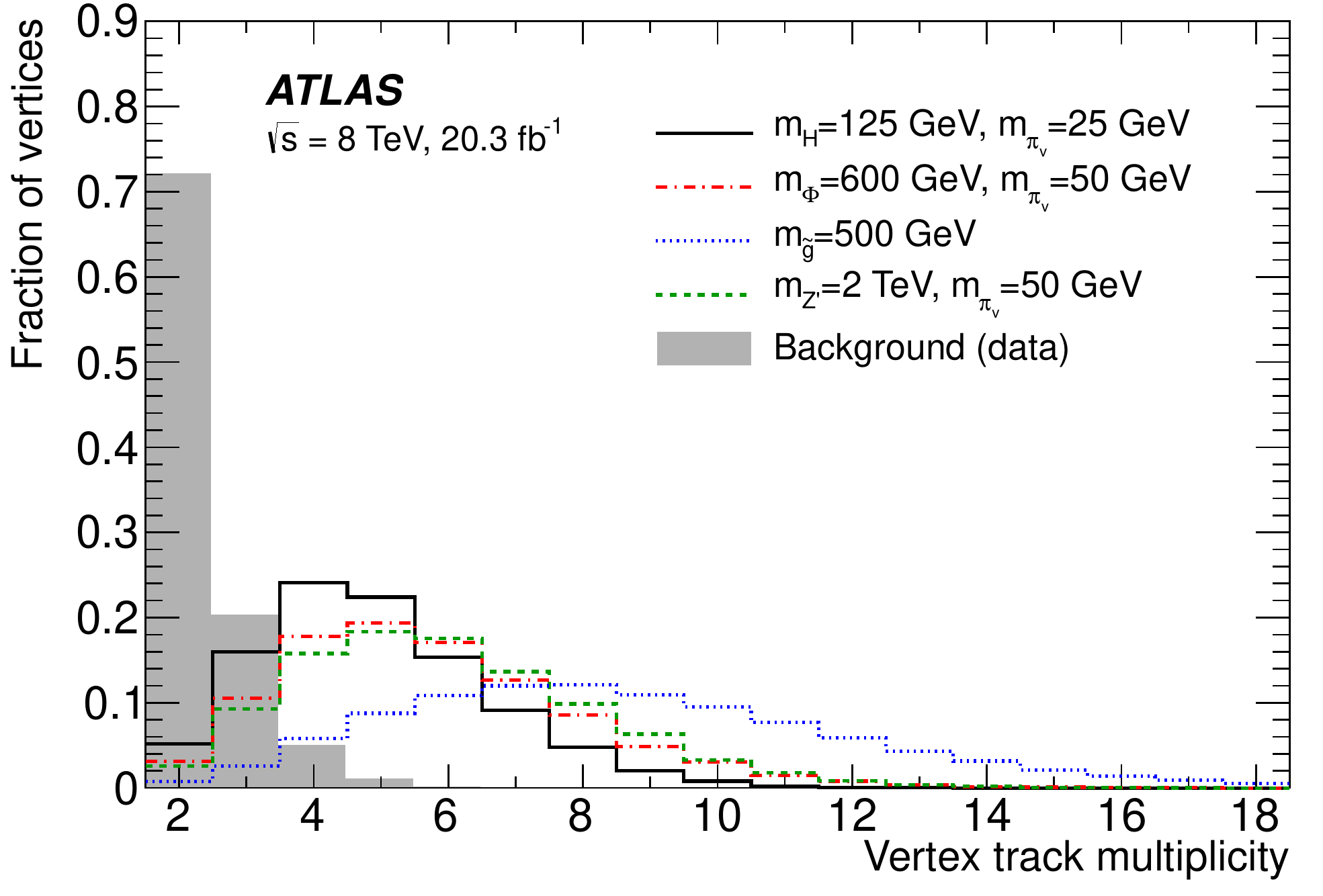}
\caption{Track multiplicity for vertices obtained from four signal benchmark samples and background multijet events.}
\label{fig:nTrksIDVx}
\end{figure}

A third selection criterion is based on the per-vertex track multiplicity. Background vertices, both from material interactions and from random intersections of tracks, have on average a lower track multiplicity than signal vertices (Fig.~\ref{fig:nTrksIDVx}). The metric $S/\sqrt{B}$ is used to select appropriate requirements on the track multiplicity, where $S$ and $B$ are the fraction of vertices retained after applying a given requirement in signal and background, respectively. A single minimum track multiplicity cut is chosen for each analysis channel. Vertices reconstructed in the Jet+\MET channel are required to contain at least seven tracks. For the Muon Cluster channel, vertices must contain at least five tracks. This lower minimum vertex track multiplicity is chosen based on the scalar boson benchmark samples with \mpiv~=~10 and 25~GeV, which have a lower track multiplicity than the other benchmark samples.

\begin{figure}[htbp]
\includegraphics[width=0.5\textwidth]{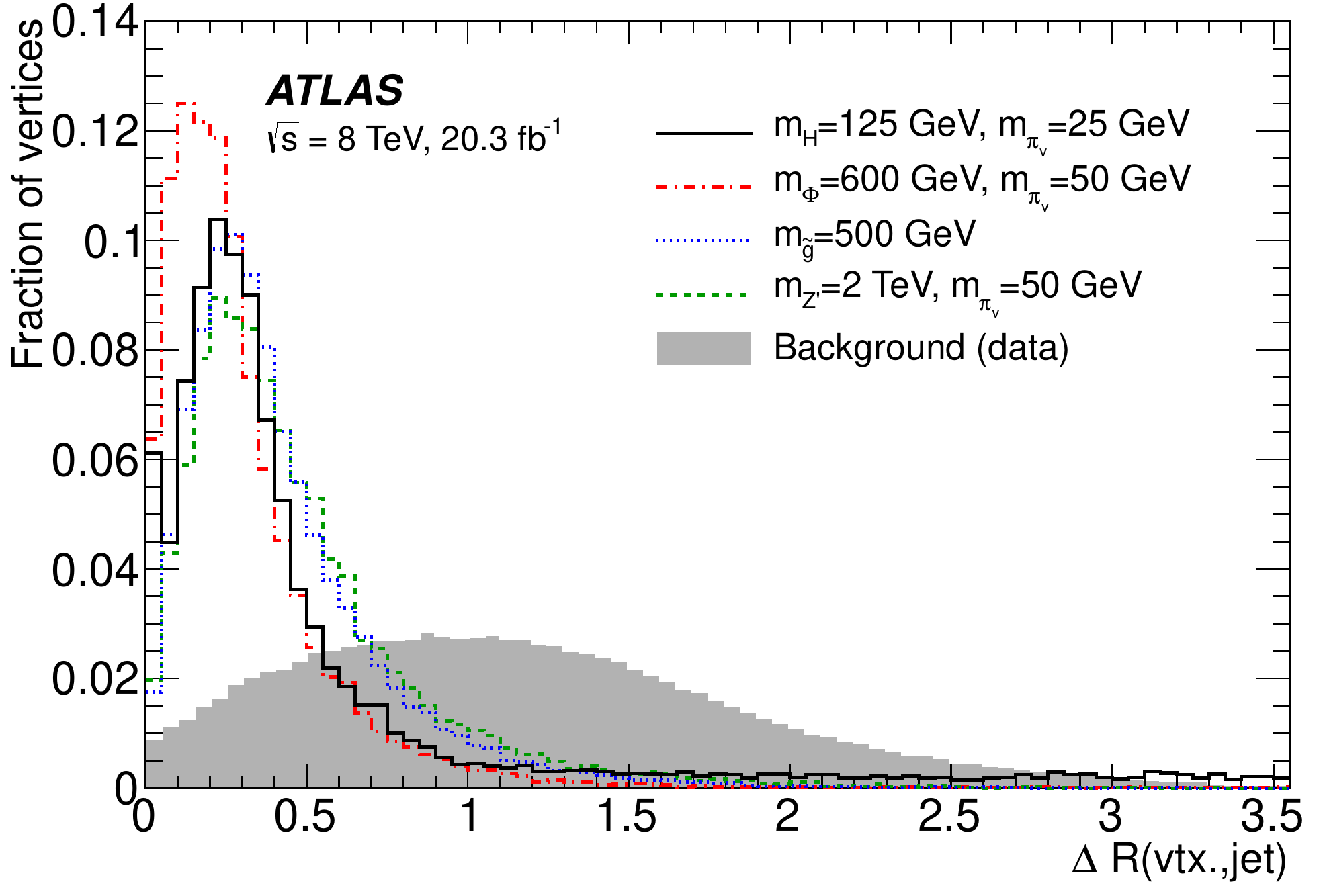}
\caption{$\Delta R$ between the ID vertex and the central axis of the nearest jet for vertices from four signal benchmark samples and background multijet events.}
\label{fig:dRIDVx}
\end{figure}

Figure~\ref{fig:dRIDVx} shows that a significant fraction of reconstructed vertices in the signal simulations are followed by a jet. The jets used in this study are required to satisfy the ATLAS medium jet criteria~\cite{JETPERF} and have $\pT>20$ GeV. The metric $S/\sqrt{B}$ is again used to select appropriate requirements, and vertices in the Jet+\MET channel must lie within a $\Delta R = 0.6$ cone around a jet axis. For the Muon Cluster channel, vertices are required to be within a $\Delta R = 0.4$ cone around a jet axis.


The final selection criteria are given in Table~\ref{tab:IDGVXsel}. The effect on the efficiency of applying the criteria to signal MC events ranges from 15--30\%, depending on the benchmark model. 

\begin{figure}[htbp]
\includegraphics[width=0.5\textwidth]{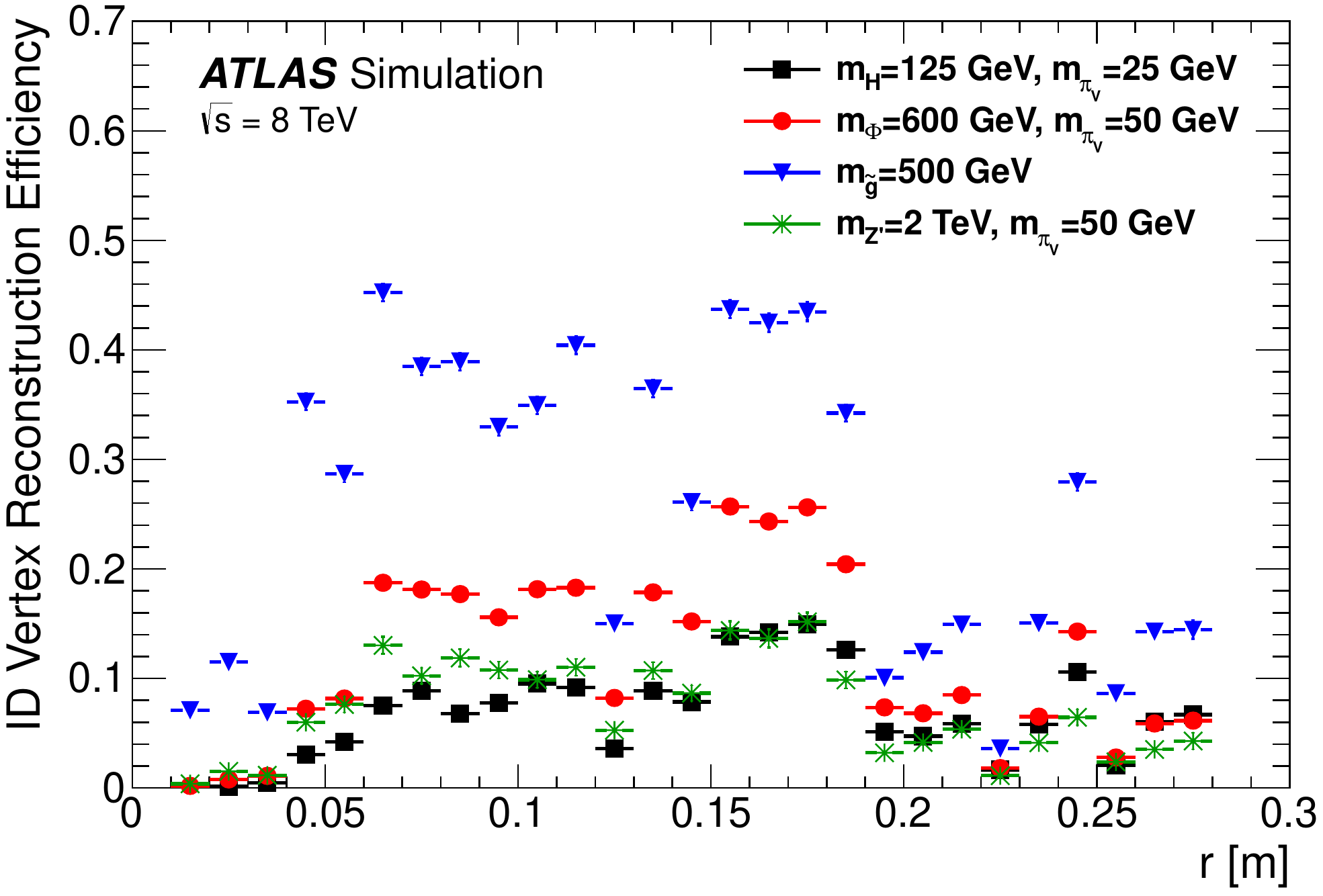}
\caption{
ID vertex reconstruction efficiency as a function of the radial decay position of the long-lived particle for scalar boson, Stealth SUSY, and \Zprime benchmark samples.
}
\label{fig:IDVx-Eff}
\end{figure}

Figure~\ref{fig:IDVx-Eff} shows the ID vertex reconstruction efficiency for a selection of benchmark samples. The efficiency is defined as the fraction of simulated decays in the ID that are matched to a reconstructed vertex satisfying the selection criteria. The fluctuations of the efficiency plot is caused by the removal of vertices that are near material.

\begin{table}[htbp]
\caption{Good-vertex criteria for vertices reconstructed in the ID.}
\begin{ruledtabular}
\begin{tabular}{lcc}
\multicolumn{1}{c}{Requirement} & \multicolumn{1}{c}{Muon Cluster channel} & \multicolumn{1}{c}{Jet+\MET channel} \\
\hline
$d/\sigma$ from material & $\geq$ 6 & $\geq$ 6 \bigstrut[t] \\
Vertex $\chi^2$ prob. &  $>$ 0.001 & $>$ 0.001 \\
$\Delta R$(vtx,jet) &$<$ 0.4 & $<$ 0.6 \\
Number of tracks & $\geq$ 5 & $\geq$ 7 \\
\end{tabular}
\end{ruledtabular}
\label{tab:IDGVXsel}
\end{table}

\subsection{Vertex reconstruction in the muon spectrometer}

Vertices in the MS are reconstructed by forming track segments from hits in the two multilayers of an MDT chamber and pairing the segments to form \textit{tracklets}. These tracklets are then fit to vertices using the algorithm detailed in Ref.~\cite{MSVXRECO}.

Detectable decay vertices are located in the region between the outer edge of the HCal and before the middle station of the muon chambers. The primary source of background to long-lived particles decaying to hadronic jets in the MS is jets that punch through the calorimeter. A series of criteria are applied to the reconstructed vertices in order to reject vertices due to background processes. These criteria were determined by comparing signal MC events to simulated multijet events generated across a broad range of jet \pT values.

The criteria are established by optimizing $S/\sqrt{B}$, where $S$ and $B$ are respectively the fraction of signal and background events that survive after imposing a particular selection criterion. Multijet events that contain vertices should have ID tracks and jets that both connect the vertex to the IP.  To reduce the acceptance of fake vertices from multijet events, good vertices are required to be isolated with respect to ID tracks and calorimeter jets. The jets considered for isolation must satisfy both  $E_{\text{T}}>30$ GeV and $\LogCaloRatio<0.5$. The value $\LogCaloRatio$ gives a measure of the fraction of energy in the jet that is deposited in the HCal ($E_{\rm{HAD}}$) and the fraction deposited in the ECal ($E_{\rm{EM}}$). This requirement ensures that vertices origimating from long-lived particles that decay near the outer edge of the hadronic calorimeter and also have significant MS activity are not rejected.

An MS vertex due to a displaced decay typically has many more hits than an MS vertex from a jet that punches through the calorimeter, so a minimum number of MDT and RPC/TGC hits is required. A maximum number of MDT hits is also applied to remove background events caused by coherent noise bursts in the MDT chambers. The minimum required number of RPC/TGC hits also helps to further reject these noisy events, since a noise burst in the MDT system is not expected to be coherent with one in the muon trigger system.  

\begin{table}
\caption{Summary of criteria for good MS vertices in the barrel and endcap regions.}
\begin{ruledtabular}
\begin{tabular}{lcc}
\multicolumn{1}{c}{Requirement} & Barrel & Endcap \\
\hline
MDT hits & 300 $\le n_{\rm{MDT}} <$ 3000 & 300 $\le n_{\rm{MDT}} <$ 3000 \bigstrut[t] \\
RPC/TGC hits & $n_{\rm{RPC}} \geq250$ & $n_{\rm{TGC}} \geq250$ \\
Track isolation & $\Delta R < 0.3$ & $\Delta R <$ 0.6 \\
Track $\Sigma \pT$ & $\Sigma \pT < 10$ GeV & $\Sigma \pT < 10$ GeV \\
Jet isolation & $\Delta R < 0.3$ & $\Delta R < 0.6$  \\
\end{tabular}
\end{ruledtabular}
\label{tab:vertexsel}
\end{table}

Table~\ref{tab:vertexsel} summarizes the optimized criteria for selecting good MS vertices. These criteria select about 60--70\% (40--60\%)  of MS vertices for the scalar boson and Stealth SUSY (\Zprime) samples, with a moderate difference between the various samples, while reducing the multijet background to a negligible value.

\begin{figure}[htbp]
	\includegraphics[width=0.5\textwidth]{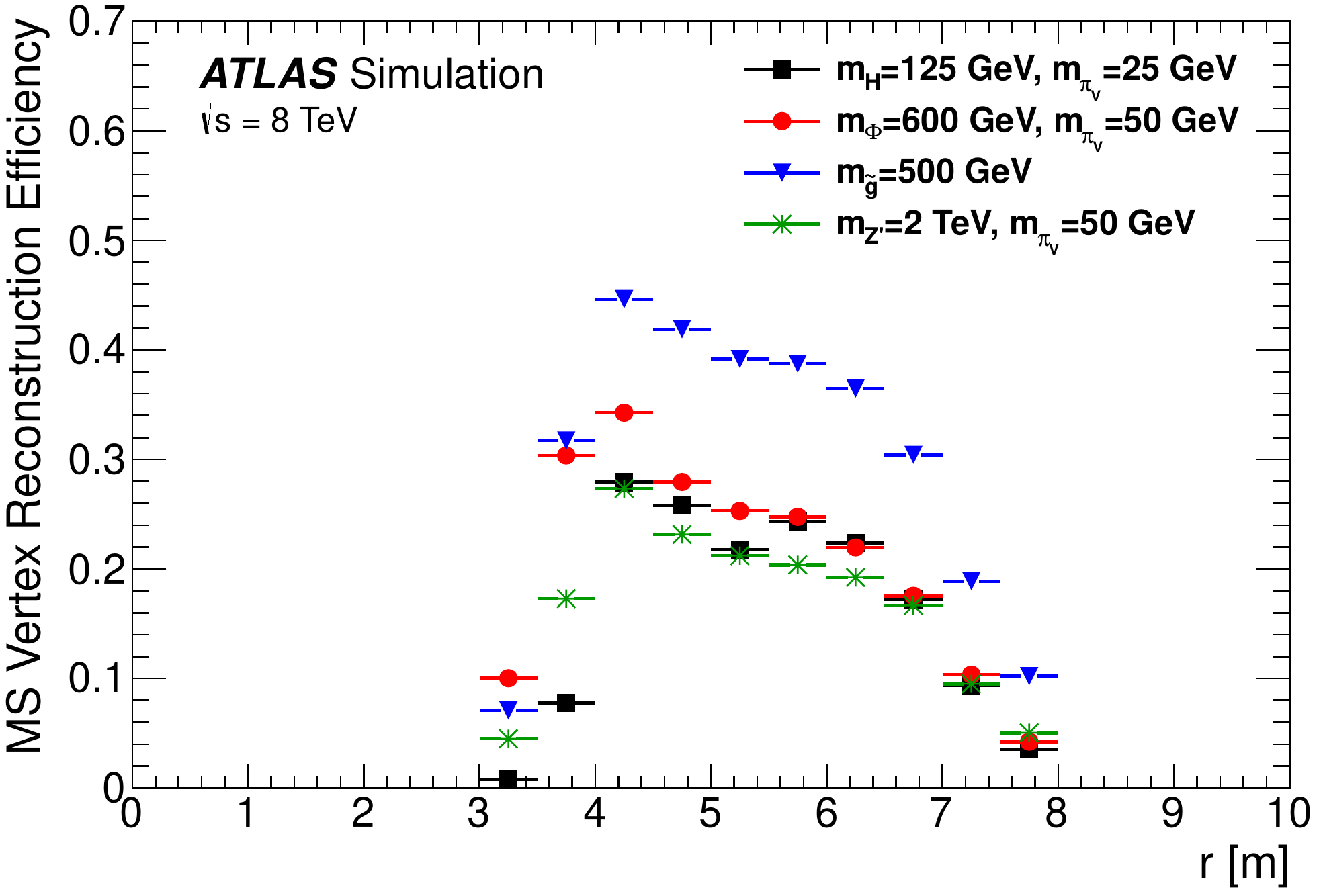}
	\caption{
	Barrel MS vertex reconstruction efficiency as a function of the radial decay position of the long-lived particle for scalar boson, Stealth SUSY, and \Zprime benchmark samples.
	}  
	\label{fig:MSVertexEffBarrel}
\end{figure}

\begin{figure}[htbp]
	\includegraphics[width=0.5\textwidth]{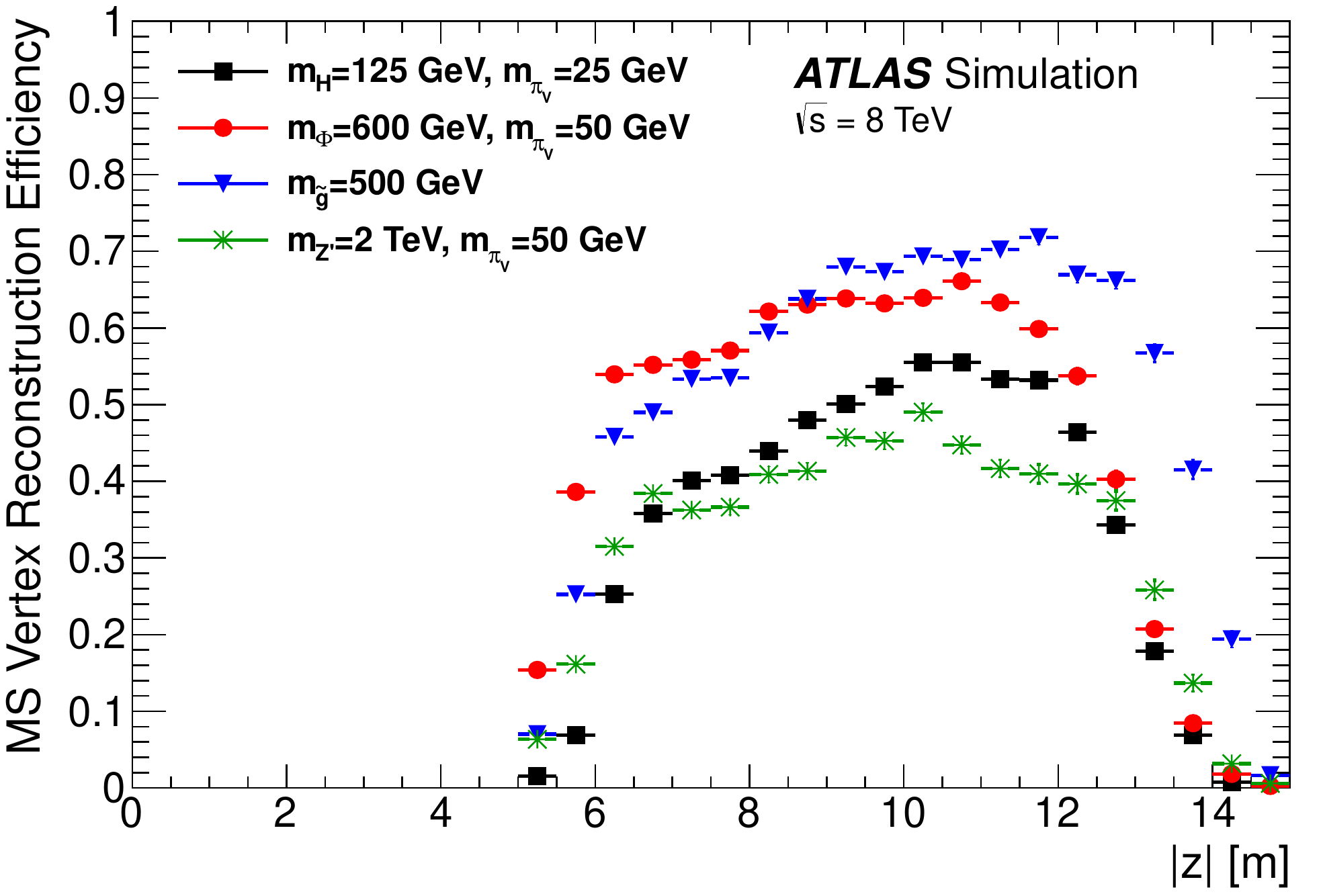}
	\caption{
	Endcap MS vertex reconstruction efficiency as a function of the $|z|$ decay position of the long-lived particle for scalar boson, Stealth SUSY, and \Zprime benchmark samples.
	}
	\label{fig:MSVertexEffEndcap}
\end{figure}

The efficiency for vertex reconstruction is defined as the fraction of simulated long-lived particle decays in the MS fiducial volume that match a reconstructed vertex satisfying all of the good-vertex criteria. A vertex is considered matched to a displaced decay if the vertex is within $\Delta R = 0.4$  of the simulated decay position. Figure~\ref{fig:MSVertexEffBarrel} shows the efficiency for reconstructing a vertex in the MS barrel for a selection of benchmark samples. Figure~\ref{fig:MSVertexEffEndcap} shows the efficiency for reconstructing a vertex in the MS endcaps.

  
The MS barrel vertex reconstruction efficiency is 30--40\% near the outer edge of the hadronic calorimeter ($r\approx4$\,m) and it substantially decreases as the decay occurs closer to the middle station ($r\approx7$\,m). The decrease occurs because the charged hadrons and photons are not spatially separated and overlap when they traverse the middle station. This results in a reduction of the efficiencies for track reconstruction and, consequently, vertex reconstruction. The efficiency for reconstructing vertices in the MS endcaps reaches 70\% for higher-mass benchmark models. Because there is no magnetic field in the region in which endcap tracklets are reconstructed, the vertex reconstruction algorithm does not have the same constraints that are present in the barrel. Consequently, the vertex reconstruction in the endcaps is more efficient for signal, but also less robust in rejecting background events. Details are provided in Ref.~\cite{MSVXRECO}.

\section{Systematic uncertainties} \label{sec:systematics}
\subsection{ID vertex reconstruction}

A multijet control sample is used to understand possible differences in track and vertex reconstruction efficiencies between data and simulated samples. The ID vertex reconstruction uncertainty is determined in two steps. 

The first step is to determine the difference in reconstruction efficiency for barrel and endcap tracks with impact parameters larger than those allowed by default track reconstruction. The distribution of reconstructed vertex positions from $K^{0}_{S}$ decays is determined for both the data and simulated multijet samples, and the distributions are normalized such that the same number of $K^{0}_{S}$ vertices are reconstructed inside the beam pipe. The weighted averages of the data-to-MC ratios of the normalized decay position distributions are 0.99~$\pm$~0.03 in the barrel and 1.01~$\pm$~0.05 in the endcaps.

The statistical uncertainties are taken as the systematic uncertainties on the $K^{0}_{S}$ reconstruction efficiency. Since two tracks are necessary for reconstructing a $K^{0}_{S}$ decay, the $K^{0}_{S}$ reconstruction efficiency is proportional to the square of the track efficiency, from which per-track systematic uncertainties of 2\% and 3\% in the barrel and endcaps, respectively, are derived. 

The second step randomly removes 2\% of barrel tracks and 3\% of endcap tracks and reruns the ID vertex reconstruction algorithm. The results provide a measure of the systematic uncertainty on ID vertex reconstruction.

The simulated samples contain pileup interactions as recorded with zero-bias triggers for a portion of the data. A reweighting to match the $\langle\mu\rangle$ distribution\footnote{The quantity $\mu$ is a measure of the average number of inelastic interactions per bunch crossing and $\langle\mu\rangle$, the average value over all proton bunches, gives the average number of expected proton-proton collisions per event.} in the full 2012 dataset is applied to take into account differences in the instantaneous luminosity between the full 2012 dataset and the sample used for zero-bias overlay. A systematic uncertainty due to pileup is determined by shifting the $\langle\mu\rangle$ distribution from MC simulation up and down by the statistical uncertainty. 

Uncertainties on the jet energy scale (JES)~\cite{JETPERF2} and PDFs are also considered and provide a small contribution to the total systematic uncertainty. Other possible sources of systematic uncertainty, such as initial- or final-state radiation, are found to be negligible. A summary of the total ID vertex reconstruction systematic uncertainties for signal MC events is presented in Table~\ref{tab:SystUncertainty}. 

\begin{table}[ht!]
\caption{Summary of the systematic uncertainties on displaced vertex reconstruction efficiencies in the ID and MS.}
\begin{ruledtabular}
\begin{tabular}{cccD{.}{.}{1,1}D{.}{.}{1,1}D{.}{.}{2,1}}
\multirow{2}{*}{$m_{\varPhi}$ [GeV]} & \multirow{2}{*}{\mpiv [GeV]} & \multirow{2}{*}{IDVx [\%]} & \multicolumn{2}{c}{MSVx [\%]} \\
\cline{4-5}
& & &  \multicolumn{1}{c}{barrel} &  \multicolumn{1}{c}{endcaps} \bigstrut \\
\hline
100 & 10  & 2.7  &  6.8  & 11.2 \bigstrut[t] \\
100 & 25  & 2.1  &  6.4  & 10.4  \\
125 & 10  & 2.5  &  7.0  & 9.9  \\
125 & 25  & 2.5  &  6.8  & 9.7  \\
125 & 40  & 2.4  &  6.5  & 8.0  \\
140 & 10  & 2.7  &  7.0  & 9.6  \\
140 & 20  & 2.7  &  6.6  & 9.6  \\
140 & 40  & 1.6  &  6.6 &  7.9  \\
300 & 50  & 2.7  &  6.9  & 6.3  \\
600 & 50  & 2.9  &  6.8  & 5.4  \\
600 & 150& 3.1  &  6.6  & 4.0  \\
900 & 50  & 3.5  &  6.6  & 5.7  \\
900 & 150& 3.0  &  5.9  & 3.8 \bigstrut[b] \\
\hline\hline
\multicolumn{2}{c}{\multirow{2}{*}{$m_{\tilde{g}}$ [GeV]}} & \multirow{2}{*}{IDVx [\%]} & \multicolumn{2}{c}{MSVx [\%]} \bigstrut \\
\cline{4-5}
& & &  \multicolumn{1}{c}{barrel} &  \multicolumn{1}{c}{endcaps} \bigstrut \\
\hline
\multicolumn{2}{c}{110}   & 3.8  &  5.6  & 4.0 \bigstrut[t] \\
\multicolumn{2}{c}{250}   & 2.3  &  5.8  & 3.8  \\
\multicolumn{2}{c}{500}   & 2.4  &  6.3  & 3.8  \\
\multicolumn{2}{c}{800}   & 2.7  &  6.5  & 3.5  \\
\multicolumn{2}{c}{1200} & 1.5  &  6.6  & 3.8 \bigstrut[b] \\
\hline\hline
\multirow{2}{*}{$m_{\Zprime}$ [TeV]} & \multirow{2}{*}{\mpiv [GeV]} & \multirow{2}{*}{IDVx [\%]} & \multicolumn{2}{c}{MSVx [\%]} \bigstrut \\
\cline{4-5}
& & &  \multicolumn{1}{c}{barrel} &  \multicolumn{1}{c}{endcaps} \bigstrut \\
\hline
1 & 50  & 2.5  &  6.8  & 6.3 \bigstrut[t] \\
2 & 50  & 2.6  &  7.0  & 6.6  \\
2 & 120 & 2.2  &  6.6  & 5.2\\
\end{tabular}
\end{ruledtabular}
\label{tab:SystUncertainty}
\end{table}

\subsection{MS trigger and vertex reconstruction}

Jets that punch through the calorimeter and shower in the MS are used to evaluate MS-related systematic uncertainties due to imperfect modeling in the simulation. Punch-through jets are similar to signal events as they contain low-energy photons as well as charged hadrons in a localized region of the MS. A punch-through jet is selected by requiring the jet to be located in the regions with reduced total interaction length, $0.7 < | \eta | < 1.0$ or $1.5 < | \eta| < 1.7$; have $| \Delta\phi(\MET,\rm{jet}) | < 0.6$ in addition to $\MET>30\GeV$; and have at least 250 MDT hits within a cone of $\Delta R < 0.6$ around the jet axis. 

\begin{figure}[htbp]
\includegraphics[width=0.5\textwidth]{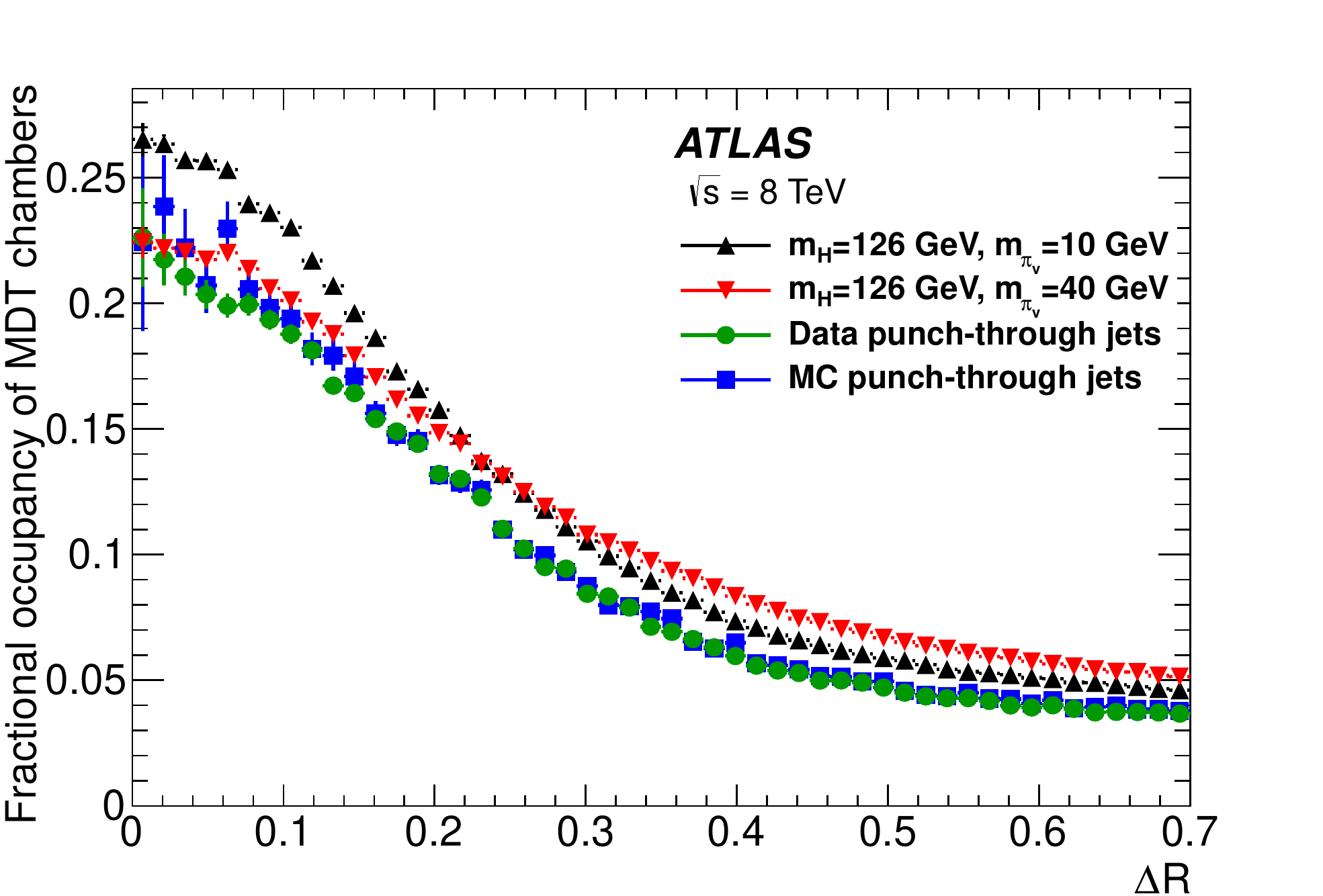}
\caption{Average fractional occupancy of chambers as a function of $\Delta R$ between the center of the MDT chamber and either the axis of the punch-through jet for the multijet samples or the \piv decay position for signal MC samples. Only punch-through jets with at least 250 MDT hits are included.}
\label{fig:avgOccupancy}
\end{figure}

The fractional occupancy of a given chamber is defined as the number of observed MDT hits divided by the total number of MDT tubes. Figure~\ref{fig:avgOccupancy} gives the average fractional occupancy of chambers as a function of $\Delta R$ between the center of the MDT chamber and either the  for the multijet samples or the \piv decay position for two scalar boson benchmark models.

The peak occupancy for both MC signal samples and the punch-through jet data and MC samples is about $20-25$\%. The punch-through jet distribution has a similar shape to that from signal events, and thus punch-through jets are a reasonable sample for studying systematic uncertainties due to incorrect simulation modeling.

\subsubsection{Muon RoI Cluster trigger}
In order to determine any differences in muon RoI reconstruction between data and simulation, the number of muon RoIs within the punch-through jet cone ($\Delta R < 0.6$) is determined for both the data and simulated samples. The weighted average of the ratio of these plots is found to be 0.91~$\pm$~0.06 in the barrel and 0.99~$\pm$~0.06 in the endcaps. The barrel value differs from unity by more than one standard deviation, so the 9\% deviation is taken as the systematic uncertainty. The value in the endcaps is consistent with unity within one standard deviation, so the 6\% statistical uncertainty is taken as the systematic uncertainty.

To understand the effects of the JES uncertainty on the isolation requirement, the energy of each jet is increased by its JES uncertainty. The resulting change in trigger efficiency is less than 0.1\% for signal MC samples. The systematic uncertainty due to pileup uncertainties is less than 0.3\% for every signal MC sample. 
Uncertainties in the initial-state radiation (ISR) spectrum~\cite{ISR} have a negligible effect on trigger acceptance.

The total systematic uncertainties of the Muon RoI Cluster trigger are 9\% in the barrel and 6\% in the endcaps, and are dominated by the data--MC simulation comparison of punch-through jets. 

\subsubsection{MS vertex reconstruction}

The systematic uncertainty on the MS vertex reconstruction efficiency due to discrepancies between data and simulation is determined using a similar method to the one for determining the systematic uncertainty on ID vertex reconstruction. First, a comparison between data and MC simulation is made using the distributions of the number of tracklets found within a $\Delta R < 0.6$ cone of a punch-through jet. The weighted averages of the ratios are 0.96~$\pm$~0.05 in the barrel and 0.89~$\pm$~0.05 in the endcaps, from which systematic uncertainties for tracklet finding of 5\% and 11\% are assigned to the barrel and endcaps, respectively. The systematic uncertainty of tracklet reconstruction is propagated to MS vertex reconstruction by randomly removing tracklets with a probability equal to the systematic uncertainty on the tracklet reconstruction efficiency.  The change in the efficiency to reconstruct MS vertices gives the systematic uncertainty for each MC signal sample. 

To understand the effects of the JES, pileup, PDF, and ISR uncertainties on MS vertex reconstruction, the same procedure used for the Muon RoI Cluster trigger and ID vertex reconstruction is applied. All of the resulting systematic uncertainties associated with these effects are less than 1\%. The total systematic uncertainty of reconstructing a vertex is found by adding in quadrature the contributions from all of the above systematic uncertainties, and the results are summarized in Table~\ref{tab:SystUncertainty}. As with the Muon RoI Cluster trigger, the dominant contribution to the uncertainty is from the MC accuracy, evaluated using a comparison of punch-through jets in data and MC.

\section{Expected number of background events} \label{sec:expected-background}
\subsection{ID vertex fake rate}
 
Jets with high track multiplicity are a major source of fake ID vertices. The ID vertex fake rate is calculated from jets in events passing single-jet triggers using three control regions.

For the \MuonClusterChannel, the control region contains events with leading jet $\pT>230\GeV$.  A per-jet fake rate as a function of \pT is derived by calculating the probability of finding a good ID vertex in the cone formed by the nonleading jets from the control region. The large \pT requirement on the leading jet ensures minimal signal contamination, and selecting only nonleading jets provides a jet-\pT spectrum comparable to the signal region. The per-jet fake rate ranges from $2\times10^{-5}$ to $3\times10^{-4}$, with uncertainties of about 20\%.

A slightly different procedure is employed for the \JetChannel. Here, two control regions are defined, both with $\MET<75\GeV$ in order to minimize signal contamination. The first contains events with a leading jet with $120\GeV<\pT<300\GeV$ and the second with a leading jet $\pT>300\GeV$. A fake rate as a function of \pT is obtained separately for leading and nonleading jets. Since no nonleading jets associated with background vertices are present in the low-\pT control region, a $\text{scale factor}$ defined as $(\text{nonleading-jet fake rate})/(\text{leading-jet fake rate})$ is computed in the high-\pT control region. This scale factor is multiplied by the leading-jet fake rate in the low-\pT control region, which gives an estimate of the nonleading-jet fake rate for the low-\pT control region. The per-jet fake rate from nonleading jets ranges from $6\times10^{-6}$ to $3\times10^{-5}$, while for leading jets the range is from $4\times10^{-6}$ to $2\times10^{-5}$, again with uncertainties of about 20\%.

\subsection{MS vertex fake rate}
Events with a single MS vertex that pass either the Muon RoI Cluster trigger or a set of minimum-bias triggers are used to determine the MS vertex fake rate. These events are dominated by fake vertices that do not originate from displaced decays. Signal events could contaminate this control region, which would result in overestimation of the probabilities and resulting background rates. 

The probability of finding a fake vertex in events not selected by the Muon RoI Cluster trigger, \PMSVertex, is determined from data selected by minimum-bias triggers. This sample is used to determine the probability that, in a random event, an MS vertex not associated with a cluster of muon RoIs is reconstructed. Such a vertex could be caused by a jet that passes the isolation criteria and showers in the MS, detector noise, cavern background, or cosmic rays. The fake rate is calculated as the number of good MS vertices divided by the total number of events. No good MS vertices are observed. Therefore zero is taken as the central value, and an upper systematic uncertainty of three events is computed as the 95\% confidence level (CL) upper bound corresponding to zero observed events $(0^{+3}_{-0})$. Dividing this by the number of events in the sample gives \PMSVertex \ equal to $(0^{+5}_{-0})\times 10^{-7}$.

The probability of finding a vertex given a muon RoI cluster is needed to find the expected background from events with two muon RoI clusters. Two probabilities are calculated: one for finding an MS vertex given a muon RoI cluster in the barrel, \PMSRecoBarrel, and another for finding an MS vertex given a muon RoI cluster in the endcaps, \PMSRecoEndcaps. These are calculated by counting the number of events with one MS vertex and one muon RoI cluster and dividing by the number of events with one muon RoI cluster. The resulting probabilities are shown in Table~\ref{tab:backgroundProbs}.

\begin{table}[h!]
\caption{The probabilities needed to compute the background from events containing fake MS vertices.  \PMSRecoBarrel \ and \PMSRecoEndcaps \ are the probabilities of finding one MS vertex in events passing the Muon RoI Cluster trigger that have one muon RoI cluster in the barrel or endcaps, respectively. The probability of finding an MS vertex not matched to a cluster of muon RoIs is given by \PMSVertex.}
\begin{ruledtabular}
\begin{tabular}{lcc}
\multicolumn{1}{c}{Quantity} & Value \\
\hline
\PMSVertex & (0\,$^{+\,5}_{-\,0}) \times10^{-7}$ \bigstrut[t] \\
\PMSRecoBarrel & $(1.289 \pm 0.006) \times 10^{-2}$ \bigstrut \\
\PMSRecoEndcaps & $(8.00 \pm 0.03) \times 10^{-2}$ \\
\end{tabular}
\end{ruledtabular}
\label{tab:backgroundProbs}
\end{table}

\subsection{Predicted number of background events}

The number of background events is estimated for each of the topologies and triggers listed in Table~\ref{tab:topo}. 

For the \MuonClusterChannel, the number of events with a combination of an MS and ID vertex is estimated by multiplying the number of events that pass the Muon RoI Cluster trigger and have an MS vertex by the probability of finding a fake ID vertex in the event. For each event, the probability of a fake ID vertex being found is computed by multiplying the number of jets at a given \pT by the ID vertex per-jet fake rate for jets of that \pT. The resulting expected number of events is 2.0~$\pm$~0.4. 

The expected number of background events with two MS vertices is calculated by considering two contributions. The events selected by the Muon RoI Cluster trigger and containing only one MS vertex are separated into those containing two muon RoI clusters, where only one cluster is matched to the reconstructed MS vertex and the other is unmatched in the barrel or endcaps (\nEventsTwoMSTrigOneUMBarrel, \nEventsTwoMSTrigOneUMEndcaps), and those containing only one cluster of muon RoIs (\nEventsMSTrig). These numbers of events are listed in  Table~\ref{tab:backgroundNums}. The probabilities of finding an MS vertex whether or not the vertex is matched to a muon RoI cluster (\PMSRecoBarrel, \PMSRecoEndcaps,\ or \PMSVertex) are then used to calculate the expected number of fake vertices for each case, respectively. These values yield an expected number of background 2MSVx events of $0.4^{+0.3}_{-0.2}$. 

\begin{table}[h!]
\caption{The numbers of events necessary to calculate the background prediction. \nEventsMSTrig, \nEventsTwoMSTrigOneUMBarrel, and \nEventsTwoMSTrigOneUMEndcaps \ are the numbers of events with only one MS vertex containing either one muon RoI cluster (1cl) or two muon RoI custers (2cl) with one matched to the MS vertex and a second unmatched and located in the barrel/endcaps (1UMBcl/1UMEcl). \nEventsJMETTrig \ is the number of events that pass the Jet+\MET trigger and contain one MS vertex.}
\begin{ruledtabular}
\begin{tabular}{lcc}
\multicolumn{1}{c}{Quantity} & Value \\
\hline
\nEventsMSTrig & $(1.04 \pm 0.03) \times 10^{5}$ \bigstrut[t] \\
\nEventsTwoMSTrigOneUMBarrel & 9\,$^{+\,4}_{-\,3}$ \bigstrut \\
\nEventsTwoMSTrigOneUMEndcaps & 4\,$^{+\,3}_{-\,2}$ \bigstrut \\
\nEventsJMETTrig & $29 \pm 5$    \\
\end{tabular}
\end{ruledtabular}
\label{tab:backgroundNums}
\end{table}

For the \JetChannel, the expected backgrounds are calculated using slightly different inputs. To estimate the number of background 2IDVx events, the per-jet ID vertex fake rate is applied to events that pass the Jet+\MET trigger and contain at least two jets. The expected number of background events for this topology is \mbox{$(1.8\pm0.4)\times10^{-4}$}. The number of background IDVx+MSVx events is estimated by selecting events with an MS vertex, and then calculating the expected number of events with a fake ID vertex given the jets present in each event. The expected number of events obtained is \mbox{$(5.5\pm1.4)\times10^{-4}$}. 

To estimate the number of background 2MSVx events, the number of events passing event selection with an MS vertex, \nEventsJMETTrig, is multiplied by the probability of finding a fake MS vertex in a random event, \PMSVertex. The prediction for this topology is \mbox{$(0.0^{+1.4}_{-0.0})\times10^{-5}$} events.

Table \ref{tab:BackgroundPred} summarizes the number of expected events from background sources. The uncertainties quoted in the table are statistical uncertainties; systematic uncertainties are negligible.

\begin{table}[hbtp]
\caption{Number of events predicted for different final-state topologies.}
\begin{ruledtabular}
\begin{tabular}{llc}
\multicolumn{1}{c}{Trigger} & \multicolumn{1}{c}{Topology} & Predicted \\
\hline
Jet+$\MET$ & 2IDVx & $(1.8 \pm 0.4) \times 10^{-4}$ \bigstrut[t] \\
Jet+$\MET$ & IDVx+MSVx & $(5.5 \pm 1.4) \times 10^{-4}$ \\
Jet+$\MET$ & 2MSVx & (0.0\,$^{+\,1.4}_{-\,0.0})\times10^{-5}$ \\
Muon RoI Cluster & IDVx+MSVx & $2.0\pm0.4$ \\
Muon RoI Cluster & 2MSVx & 0.4\,$^{+\,0.3}_{-\,0.2}$ \\
\end{tabular}
\end{ruledtabular}
\label{tab:BackgroundPred}
\end{table}

\vspace{-2.0em}
\section{Results}\label{sec:result}
\subsection{Expected number of signal events}

The number of expected signal events is extrapolated from the generated lifetime to the range of proper lifetimes ($c\tau$) between 0 and 100\,m using 2 million MC simulated events per benchmark sample. 
The overall probability of each event satisfying the selection criteria is evaluated from the efficiencies of reconstructing an event that passes the Muon RoI Cluster or Jet+\MET trigger and the efficiencies of reconstructing an ID or MS vertex from an LLP decay.  Other important criteria, such as the timing acceptance window for the barrel trigger, are also taken into account. Table~\ref{tab:expectedSignal126} shows the expected number of signal events for Higgs boson decays, at a proper lifetime of 2\,m.

The \Zprime extrapolation procedure has an additional complication due to the presence of many \piv decays per event. This leads to a maximum discrepancy of 15\% between the reconstructed and predicted trigger efficiencies calculated from the simulated MC samples. This 15\% discrepancy is applied in the extrapolation process as a systematic uncertainty on the trigger efficiency. 

\begin{table}[!hbtp]
\caption{Expected number of signal events at $19.5^{-1}$\,fb for the scalar boson benchmark model with $m_{H}=125$ GeV and BR$(H\to\piv\piv)=100\%$, at a \piv proper lifetime of $c\tau=2$~m. The SM Higgs boson cross-section for gluon-fusion production, $\sigma_{\mbox{\footnotesize SM}}=18.97$\,pb~\cite{HiggsXS}, is used to compute numbers of signal events. Also shown are the numbers of expected background and observed events. Uncertainties on expected signal events are statistical only.}
\begin{ruledtabular}
\begin{tabular}{lcD{,}{ \pm }{-1}cc}
\multicolumn{1}{c}{\multirow{2}{*}{Topology}} & \multirow{2}{*}{\parbox{0.8cm}{\mpiv [GeV]}} & \multicolumn{2}{c}{Expected events} & \multirow{2}{*}{\parbox{1.4cm}{Observed events}} \\ 
\cline{3-4}
 & & \multicolumn{1}{c}{Signal} & \multicolumn{1}{c}{Background} & \bigstrut \\
\hline
\multirow{3}{*}{IDVx+MSVx} & 10 & 1.9\,,\,1.4& \multirow{3}{*}{2.0\,$\pm 0.4$} & \multirow{3}{*}{0} \bigstrut[t] \\
 & 25 & 62\,,\,8 & &\\
 & 40 & 41\,,\,6 & & \bigstrut[b] \\
\hline
\multirow{3}{*}{2 MSVx } & 10 & 234\,,\,15& \multirow{3}{*}{0.4\,$^{+\,0.3}_{-\,0.2}$} & \multirow{3}{*}{2} \bigstrut[t] \\
 & 25 & 690\,,\,26 & &\\
 & 40 & 313\,,\,18 & & \\
\end{tabular}
\end{ruledtabular}
\label{tab:expectedSignal126}
\end{table}

\subsection{Limits}

After applying all selection criteria, two events are found in the \MuonClusterChannel 2MSVx topology. No events are observed in the Muon Cluster channel IDVx+MSVx topology. In the Jet+\MET channel, zero events are observed in each of the three topologies: 2IDVx, IDVx+MSV, and 2MSVx. The numbers of events found are compatible with the predicted numbers of background events. The $CL_s$ method~\cite{CLs} is used to derive upper limits on the production cross-section times branching ratio ($\sigma\times\mbox{BR}$). For the scalar boson benchmark samples, BR represents the branching ratio for $\varPhi \rightarrow \piv \piv$, while for the \Zprime benchmark samples BR represents the branching ratio for $\Zprime \rightarrow q_v \bar{q_v}$. The branching ratio for \piv decay to fermion pairs is assumed to be 100\%. For the Stealth SUSY benchmarks, BR represents the branching ratio for $\tilde{g} \rightarrow \tilde{S}g$. The branching ratios for $\tilde{S} \rightarrow S \tilde {G}$ and $S\rightarrow g g$ are both assumed to be 100\% in the Stealth SUSY model considered in this analysis.
 For the Higgs boson benchmarks, upper limits are set on $\sigma\times\mbox{BR}/\sigma_{\mbox{\footnotesize SM}}$, where $\sigma_{\mbox{\footnotesize SM}}$ is the SM Higgs production cross-section, 18.97~pb~\cite{HiggsXS}. 

A profile likelihood function~\cite{Cowan:2010js} is used as the test statistic and a frequentist calculator is used to generate toy data. The likelihood includes a Poisson probability term describing the total number of observed events. Systematic uncertainties are incorporated as nuisance parameters through their effect on the mean of the Poisson functions and through convolution with their assumed Gaussian distributions. The number of expected events in signal MC samples, together with the expected background, the observed collision events and all the systematic uncertainties are provided as input for computing the $CL_s$ value, which represents the probability for the given observation to be compatible with the signal-plus-background hypothesis.

\begin{figure*}[!hbtp]
\begin{center}
\begin{tabular}{cc}
  \includegraphics[width=\columnwidth]{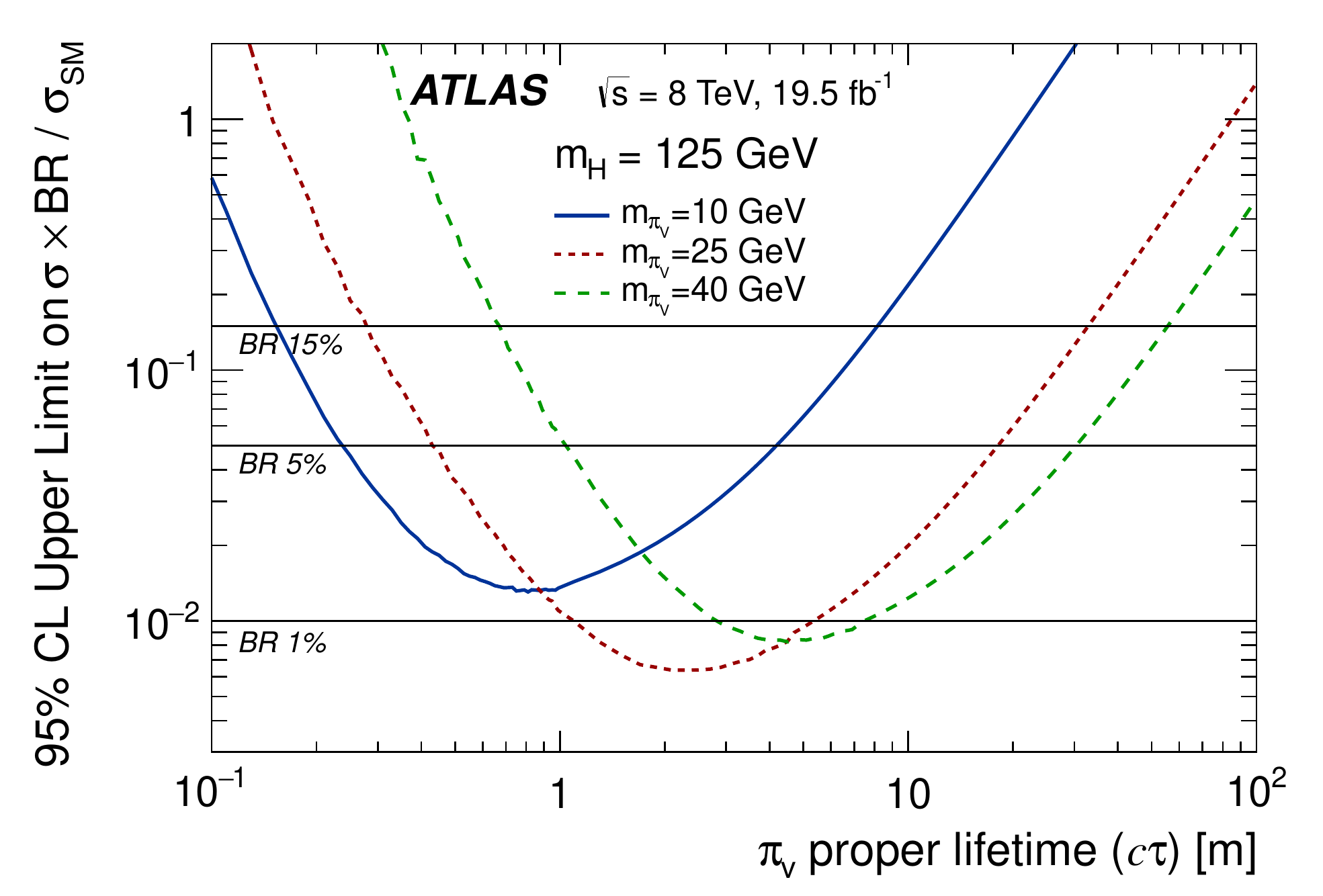} &
  \includegraphics[width=\columnwidth]{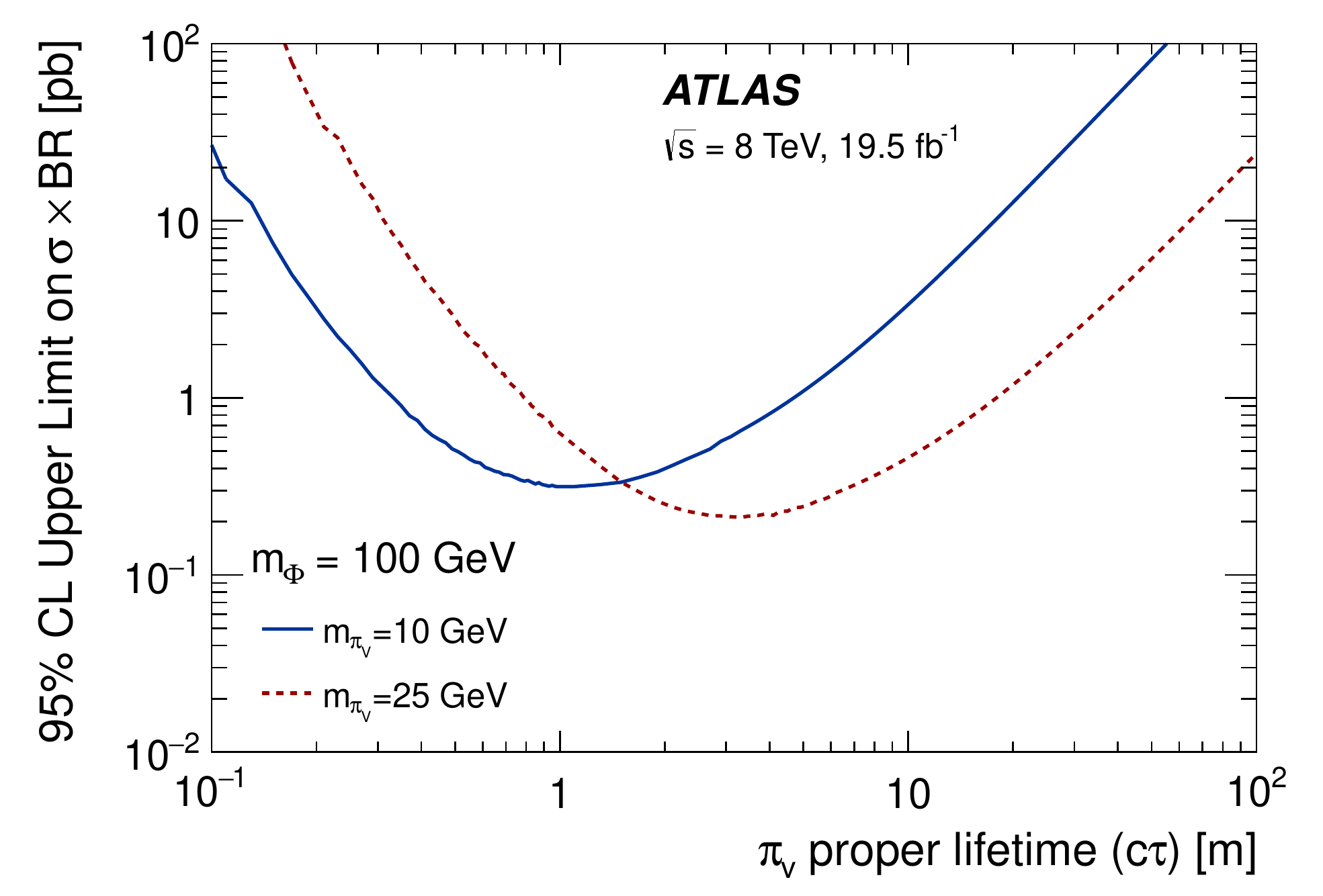}\\
  (a) & (b)\\
  \includegraphics[width=\columnwidth]{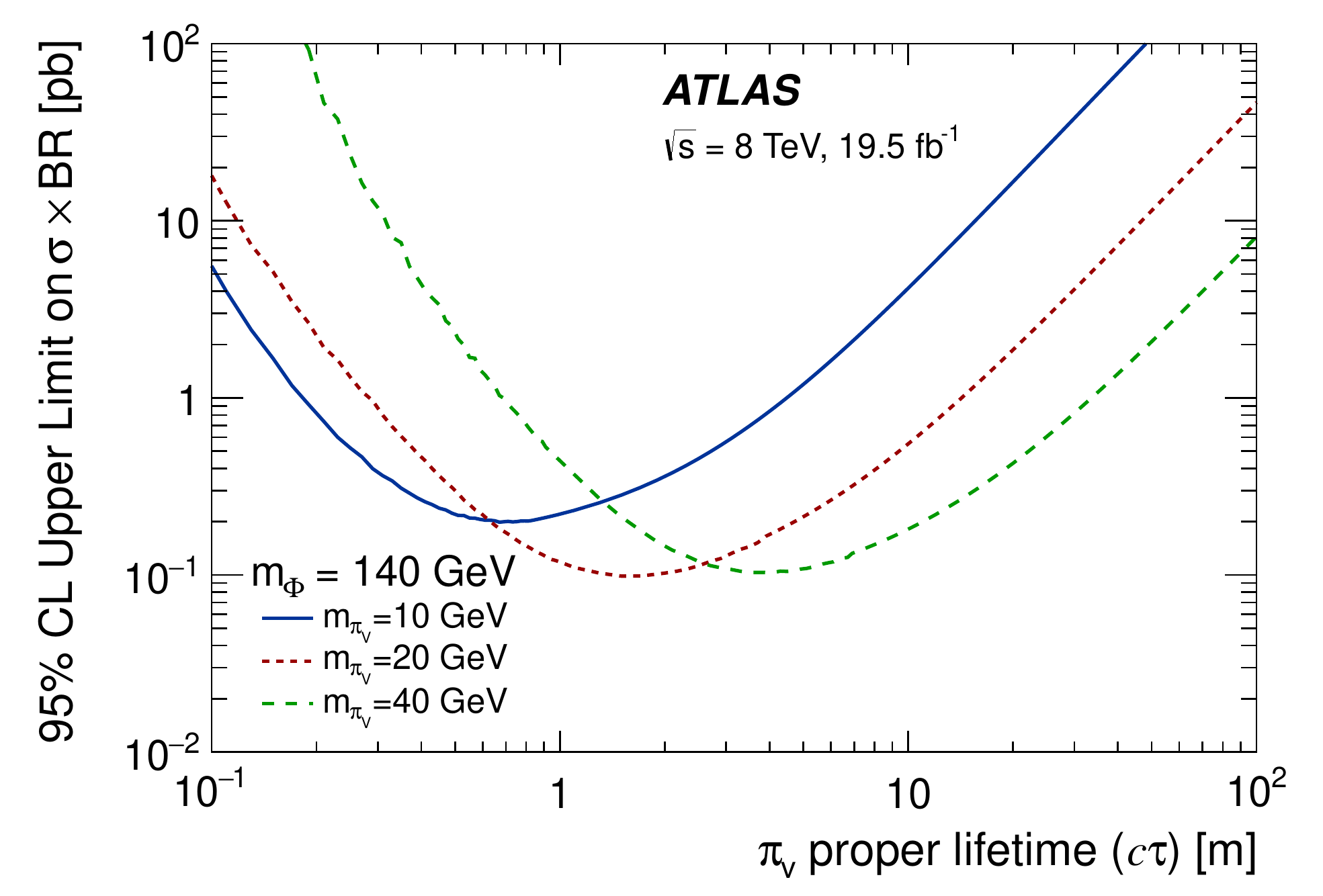} &
  \includegraphics[width=\columnwidth]{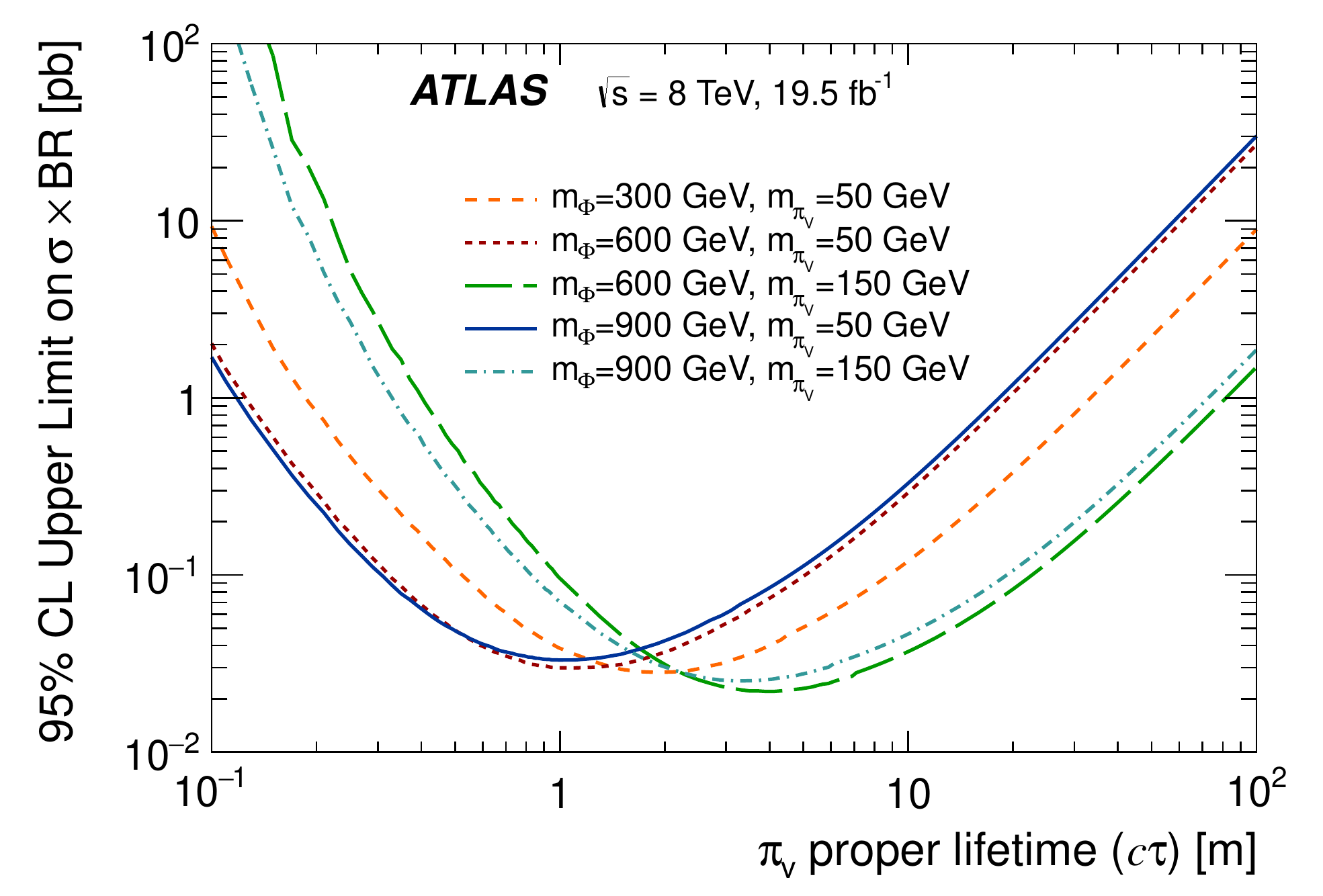}\\
  (c) & (d)\\
  \includegraphics[width=\columnwidth]{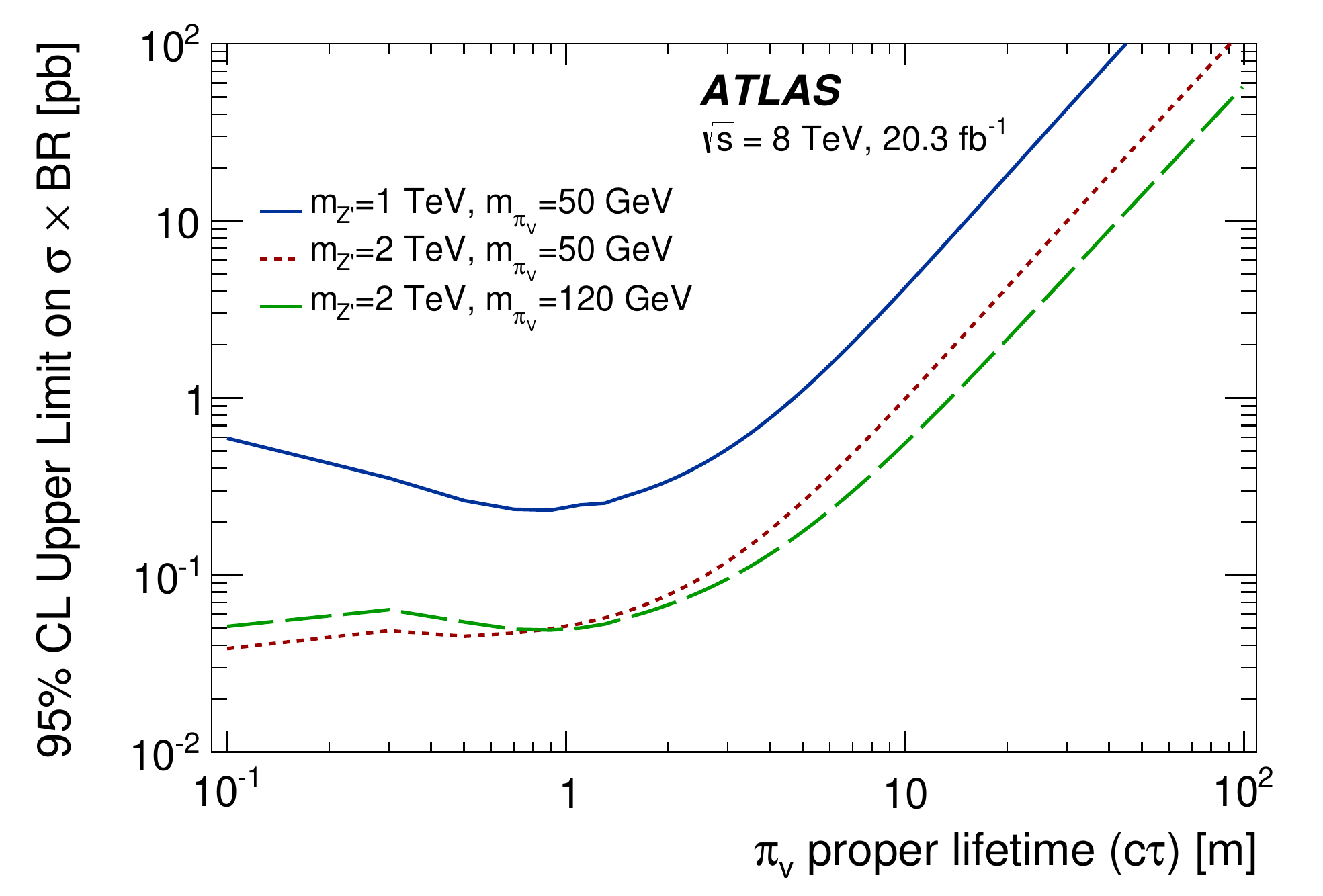} &
  \includegraphics[width=\columnwidth]{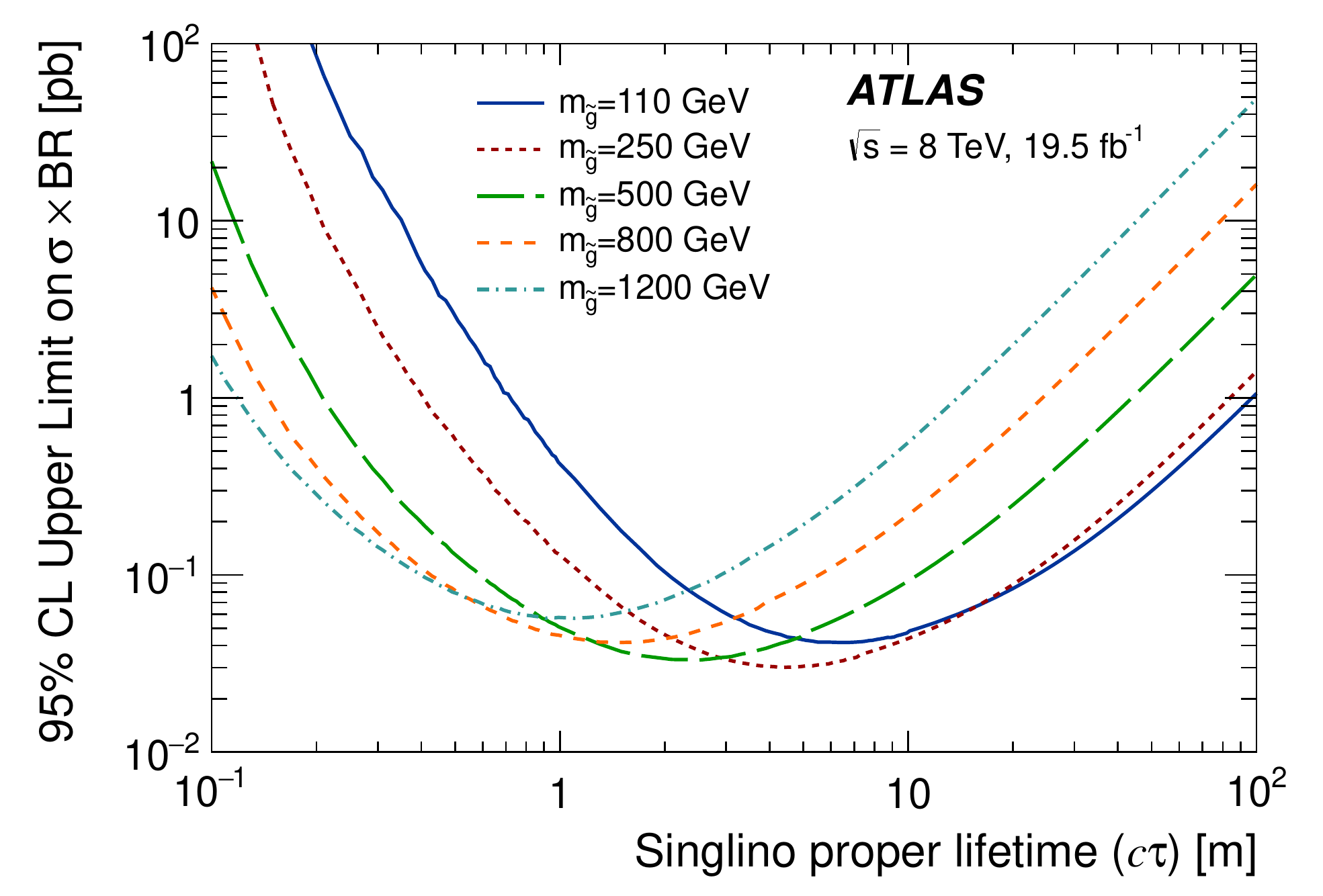}\\
  (e) & (f) \\
\end{tabular}
  \caption{\label{fig:limits} 
 (a) Observed 95\% CL limits on $\sigma\times\mbox{BR}/\sigma_{\mbox{\footnotesize SM}}$ for the scalar boson samples with $m_{H}$~=~125~\GeV. Three horizontal lines mark branching fractions for the Higgs boson decaying to \piv pairs at 15\%, 5\%, and 1\%.
  Observed 95\% CL limits on $\sigma\times\mbox{BR}$ for the scalar boson samples with 
  (b) \mPhi~=~100~\GeV,
  (c) \mPhi~=~140~\GeV,
  and (d) \mPhi~=~300~\GeV, 600~\GeV, and 900~\GeV.
  Observed 95\% CL limits on $\sigma\times\mbox{BR}$ for the (e) \Zprime samples and (f) Stealth SUSY samples.
  }
\end{center}
\end{figure*}

Observed 95\% CL upper limits are shown in Fig.~\ref{fig:limits}.  Because the SM production cross-section is known for the $m_{H}=125\GeV$ benchmark samples, excluded lifetime ranges are presented. Table~\ref{tab:limits_125} shows the excluded regions for branching ratios of 30\%, 15\%, 5\%, and 1\% for the Higgs boson decaying to long-lived particles.

\clearpage
In the \MuonClusterChannel, the 2MSVx topology provides the dominant contribution to the limits. In the Jet+\MET channel, the 2IDVx topology becomes important for probing proper lifetimes below 1~m, as can be clearly seen in Fig.~\ref{fig:limits}(e) where the limit extends to much lower proper lifetimes than in the models considered in the \MuonClusterChannel.

\begin{table}[!hbtp]
\caption{Ranges of \piv proper decay lengths excluded at 95\% CL assuming a 30\%, 15\%, 5\%, or 1\% BR for $m_{H}=125\GeV$.}
\begin{ruledtabular}
\begin{tabular}{cD{,}{-}{-1}D{,}{-}{-1}D{,}{-}{-1}D{,}{-}{-1} }
\mpiv & \multicolumn{4}{c}{Excluded $c\tau$ range [m]}  \\
\cline{2-5}
[GeV] &  \multicolumn{1}{c}{1\% BR} &  \multicolumn{1}{c}{5\% BR} &  \multicolumn{1}{c}{15\% BR} &  \multicolumn{1}{c}{30 \% BR} \bigstrut \\
\hline
10 &  \multicolumn{1}{c}{no limit} & 0.24,4.2  & 0.16,8.1 &  0.12,11.8 \bigstrut[t] \\
25 & 1.10,5.35 & 0.43,18.1 & 0.28,32.8 & 0.22,46.7  \\
40 & 2.82,7.45 & 1.04,30.4 & 0.68,55.5 &  0.52,79.2  \\
\end{tabular}
\end{ruledtabular}
\label{tab:limits_125}
\end{table}

\section{Summary}\label{sec:conclusion}
Results of a search for the decay of long-lived neutral particles to hadronic jets using proton--proton collisions at $\sqrt{s}=8$~TeV recorded at the LHC with the ATLAS detector are presented. Event selection uses two triggers: the Muon RoI Cluster trigger (19.5~fb$^{-1}$) and a Jet+\MET trigger (20.3~fb$^{-1}$). A total of five different final states with two displaced vertices are investigated, employing techniques for reconstructing highly displaced decays in the inner tracking detector and muon spectrometer.  

No significant excess of events above the background expectations is observed and exclusion limits as a function of the proper lifetime of long-lived particles from Higgs boson and scalar boson decays are reported. This paper also presents the first upper limits as a function of proper lifetime for Hidden Valley \Zprime and Stealth SUSY scenarios. For the SM Higgs boson decaying to long-lived particles, a significant range of \piv proper lifetimes is excluded for branching ratios of 5\% or less.

\section{Acknowledgments}

We thank CERN for the very successful operation of the LHC, as well as the
support staff from our institutions without whom ATLAS could not be
operated efficiently.

We acknowledge the support of ANPCyT, Argentina; YerPhI, Armenia; ARC,
Australia; BMWFW and FWF, Austria; ANAS, Azerbaijan; SSTC, Belarus; CNPq and FAPESP,
Brazil; NSERC, NRC and CFI, Canada; CERN; CONICYT, Chile; CAS, MOST and NSFC,
China; COLCIENCIAS, Colombia; MSMT CR, MPO CR and VSC CR, Czech Republic;
DNRF, DNSRC and Lundbeck Foundation, Denmark; EPLANET, ERC and NSRF, European Union;
IN2P3-CNRS, CEA-DSM/IRFU, France; GNSF, Georgia; BMBF, DFG, HGF, MPG and AvH
Foundation, Germany; GSRT and NSRF, Greece; RGC, Hong Kong SAR, China; ISF, MINERVA, GIF, I-CORE and Benoziyo Center, Israel; INFN, Italy; MEXT and JSPS, Japan; CNRST, Morocco; FOM and NWO, Netherlands; BRF and RCN, Norway; MNiSW and NCN, Poland; GRICES and FCT, Portugal; MNE/IFA, Romania; MES of Russia and NRC KI, Russian Federation; JINR; MSTD,
Serbia; MSSR, Slovakia; ARRS and MIZ\v{S}, Slovenia; DST/NRF, South Africa;
MINECO, Spain; SRC and Wallenberg Foundation, Sweden; SER, SNSF and Cantons of
Bern and Geneva, Switzerland; NSC, Taiwan; TAEK, Turkey; STFC, the Royal
Society and Leverhulme Trust, United Kingdom; DOE and NSF, United States of
America.

The crucial computing support from all WLCG partners is acknowledged
gratefully, in particular from CERN and the ATLAS Tier-1 facilities at
TRIUMF (Canada), NDGF (Denmark, Norway, Sweden), CC-IN2P3 (France),
KIT/GridKA (Germany), INFN-CNAF (Italy), NL-T1 (Netherlands), PIC (Spain),
ASGC (Taiwan), RAL (UK) and BNL (USA) and in the Tier-2 facilities
worldwide.

\providecommand{\href}[2]{#2}\begingroup\raggedright\endgroup

\clearpage

\onecolumngrid
\begin{flushleft}
{\Large The ATLAS Collaboration}

\bigskip

G.~Aad$^{\rm 85}$,
B.~Abbott$^{\rm 113}$,
J.~Abdallah$^{\rm 151}$,
O.~Abdinov$^{\rm 11}$,
R.~Aben$^{\rm 107}$,
M.~Abolins$^{\rm 90}$,
O.S.~AbouZeid$^{\rm 158}$,
H.~Abramowicz$^{\rm 153}$,
H.~Abreu$^{\rm 152}$,
R.~Abreu$^{\rm 30}$
% ...
\bigskip
\\
% ...
$^{11}$ Institute of Physics, Azerbaijan Academy of Sciences, Baku, Azerbaijan\\
% ...
$^{30}$ CERN, Geneva, Switzerland\\
% ...
$^{85}$ CPPM, Aix-Marseille Universit{\'e} and CNRS/IN2P3, Marseille, France\\
% ...
$^{90}$ Department of Physics and Astronomy, Michigan State University, East Lansing MI, United States of America\\
% ...
$^{107}$ Nikhef National Institute for Subatomic Physics and University of Amsterdam, Amsterdam, Netherlands\\
% ...
$^{113}$ Homer L. Dodge Department of Physics and Astronomy, University of Oklahoma, Norman OK, United States of America\\
% ...
$^{151}$ Institute of Physics, Academia Sinica, Taipei, Taiwan\\
$^{152}$ Department of Physics, Technion: Israel Institute of Technology, Haifa, Israel\\
$^{153}$ Raymond and Beverly Sackler School of Physics and Astronomy, Tel Aviv University, Tel Aviv, Israel\\
% ...
$^{158}$ Department of Physics, University of Toronto, Toronto ON, Canada\\
% ...
\end{flushleft}

\end{document}